\documentclass[aps,prb,twocolumn,superscriptaddress]{revtex4-2}
\usepackage{graphicx}
\usepackage[final]{changes}
\graphicspath{{./figure/}}                    
\usepackage{multirow}                 

\begin{document}


\title{Canted ferromagnetic order in a distorted triangular-lattice magnet Na$_2$SrCo(VO$_4$)$_2$}


\author{Tengfei Peng}
\affiliation{School of Physics, Beihang University, Beijing 100191, China}
\author{Xiaobai Ma}
\affiliation{Department of Nuclear Physics, China Institute of Atomic Energy, Beijing 102413, China}
\author{Xinyang Liu}
\affiliation{School of Physics, Beihang University, Beijing 100191, China}
\affiliation{Beijing National Laboratory for Condensed Matter Physics, Institute of Physics, Chinese Academy of Sciences, 100190 Beijing, China}
\author{Feiran Shen}
\affiliation{Spallation Neutron Source Science Center, Dongguan 523803, China}
\author{Lunhua He}
\affiliation{Spallation Neutron Source Science Center, Dongguan 523803, China}
\affiliation{Beijing National Laboratory for Condensed Matter Physics, Institute of Physics, Chinese Academy of Sciences, 100190 Beijing, China}
\affiliation{Songshan Lake Materials Laboratory, Dongguan 523808, China}
\author{Junsen Xiang}
\affiliation{Beijing National Laboratory for Condensed Matter Physics, Institute of Physics, Chinese Academy of Sciences, 100190 Beijing, China}
\author{Wenyun Yang}
\affiliation{State Key Laboratory for Mesoscopic Physics, School of Physics, Peking University, Beijing 100871, China}
\author{Wentao Jin}
\email{wtjin@buaa.edu.cn}
\affiliation{School of Physics, Beihang University, Beijing 100191, China}

\date{\today}

\begin{abstract}
    Triangular-lattice cobaltates with glaserite-type $X_2Y$Co($T$O$_4)_2$ structure provide an ideal platform to investigate intriguing quantum magnetism. 
    Here we report a comprehensive study of the structural and magnetic properties of a triangular-lattice cobalt vanadate $\rm Na_2SrCo(VO_4)_2$.
    Room-temperature x-ray and neutron powder diffraction confirm that $\rm Na_2SrCo(VO_4)_2$ crystallizes in the monoclinic $P2_1/c$ space group with slightly distorted triangular layers of $\rm Co^{2+}$ ions.
    Magnetization measurements reveal a ferromagnetic transition at $T\rm_C \approx 3.4~{\rm K}$, where a sharp $\lambda$-type anomaly is observed in the specific heat. 
    The magnetic entropy recovered up to 55 K approaches 90$\%$ of $R{\rm ln}2$, supporting an effective spin-1/2 state of Co$^{2+}$ ions at low temperature.
    Neutron diffraction at 2.3 K (below $T_{\rm C}$) further confirms a long-range canted ferromagnetic order with the Co$^{2+}$ moments aligned in the $ac$ plane and the ordered moment size of $\sim$ 2.6 $\mu\rm_{B}$. 
    Comparing with its sister compounds with a trigonal symmetry, $\rm Na_2BaCo(VO_4)_2$ with a collinear ferromagnetic structure and the recently discovered spin supersolid candidate $\rm Na_2BaCo(PO_4)_2$ with a distinct Y-like antiferromagnetic ground state, this study indicates the decisive role of the $T{\rm O_4}$ tetrahedra in tuning exchange interactions and contrasting magnetic behaviors of these glaserite-structure compounds.
\end{abstract}


\maketitle

\section{INTRODUCTION}
    Geometrically frustrated magnets continue to attract considerable attention, as competing interactions and lattice geometries can prevent the establishment of simple long-range magnetic order, leading to unconventional ground states and exotic excitations \cite{Starykh_2015,Wosnitza_2016,KHATUA20231,balents2010spin}.
    Among the various frustrated motifs, the triangular lattice represents one of the earliest and most extensively studied examples. 
    For spin-$\frac{1}{2}$ systems with antiferromagnetic (AFM) exchange, the triangular geometry produces strong frustration, potentially giving rise to quantum spin liquids \cite{balents2010spin,RevModPhys.89.025003,broholm2020quantum,wen2019experimental},  intrinsic quantum Ising magnets \cite{PhysRevResearch.1.033141,PhysRevResearch.2.043013}, and field-induced quantum phase transitions \cite{AVChubukov_1991,bordelon2019field,PhysRevLett.109.267206}, and so on.
    Thus, $S =\frac{1}{2}$ triangular-lattice materials have long served as canonical platforms for exploring frustrated magnetism.

    Cobalt-based triangular magnets have recently emerged as a particularly versatile class of such systems. 
    In an octahedral oxygen environment, the $\rm Co^{2+}$ ion experiences substantial spin-orbit coupling (SOC), which splits the d-manifold and often yields a Kramers doublet at low energies \cite{abragam1951theory,PhysRev.131.546,doi:10.1143/JPSJ.72.2326}. 
    This doublet can be described as an effective spin-$\frac{1}{2}$ state with a highly anisotropic $g$-tensor. 
    Consequently, cobalt-based triangular systems exhibit a broad spectrum of magnetic interactions, ranging from nearly isotropic Heisenberg-type to strongly anisotropic Ising-type couplings. 

    Extensive studies have been carried out on various layered cobaltates, in which the $\rm Co^{2+}$ ions form well-defined triangular networks separated by nonmagnetic spacer layers. 
    Representative examples include $\rm Ba_2CoTeO_6$ \cite{PhysRevB.93.094420,B927498G}, $\rm Ba_3CoSb_2O_9$ \cite{PhysRevLett.109.267206,PhysRevLett.110.267201,PhysRevLett.116.087201,ito2017structure,kamiya2018nature}, $\rm Ba_2La_2CoTe_2O_{12}$ \cite{kojima2018quantum,park2024anomalous}, $A_2{\rm Co(SeO_3)_2}$ \cite{PhysRevMaterials.4.084406,zhu2025wannier,zhu2024continuum,PhysRevB.111.L180402,shi2025absencehighfieldspinsupersolid} and $ A_2{\rm Co_2(SeO_3)_3}$ ($A$ = K, Rb) \cite{xu2024frustrated,PhysRevLett.133.136703,PhysRevB.102.224430}.
    Among these, the $\rm Na_2{\it AE}Co({\it T}O_4)_2$ ({\it AE} = Ba, Sr;{\it T} = P, V) family has drawn particular interest.
    In these compounds, triangular layers are constructed from $\rm CoO_6$ octahedra linked by nonmagnetic $T$O$_4$ tetrahedra.
    The trigonal phosphate member $\rm Na_2BaCo(PO_4)_2$ (NBCPO) with perfect triangular networks of Co exhibits antiferromagnetic interactions and hosts an intriguing Y-like spin supersolid phase below $T_{\rm N} \approx 148~{\rm mK}$, which combines the characteristics of both superfluids and solids in spin space \cite{doi:10.1073/pnas.1906483116,xiang2024giant,gao2022spin}.
    In contrast, full substitutions of $\rm Ba^{2+}$ with smaller $\rm Sr^{2+}$ ions leads to a structural transformation into the monoclinic symmetry for $\rm Na_2SrCo(PO_4)_2$ (NSCPO) \cite{zhang2022doping}, and this compound undergoes an unspecified AFM transition at $T_{\rm N} \approx 600~{\rm mK}$ \cite{PhysRevB.106.054415}. 
    Remarkably, the vanadate analogs Na$_2$$AE$Co(VO$_4$)$_2$ display drastically different behavior: several members of this family exhibit ferromagnetic (FM) or weakly canted FM order \cite{Nakayama_2013,PhysRevB.85.214422,doi:10.1021/acs.inorgchem.7b02024}. 
    This stark contrast underscores the decisive role of the nonmagnetic ligand group ($\rm PO_4$ vs. $\rm VO_4$) in determining the sign and magnitude of magnetic exchange interactions. 

\begin{figure*}
    \centering
    \includegraphics[width = 0.7\linewidth]{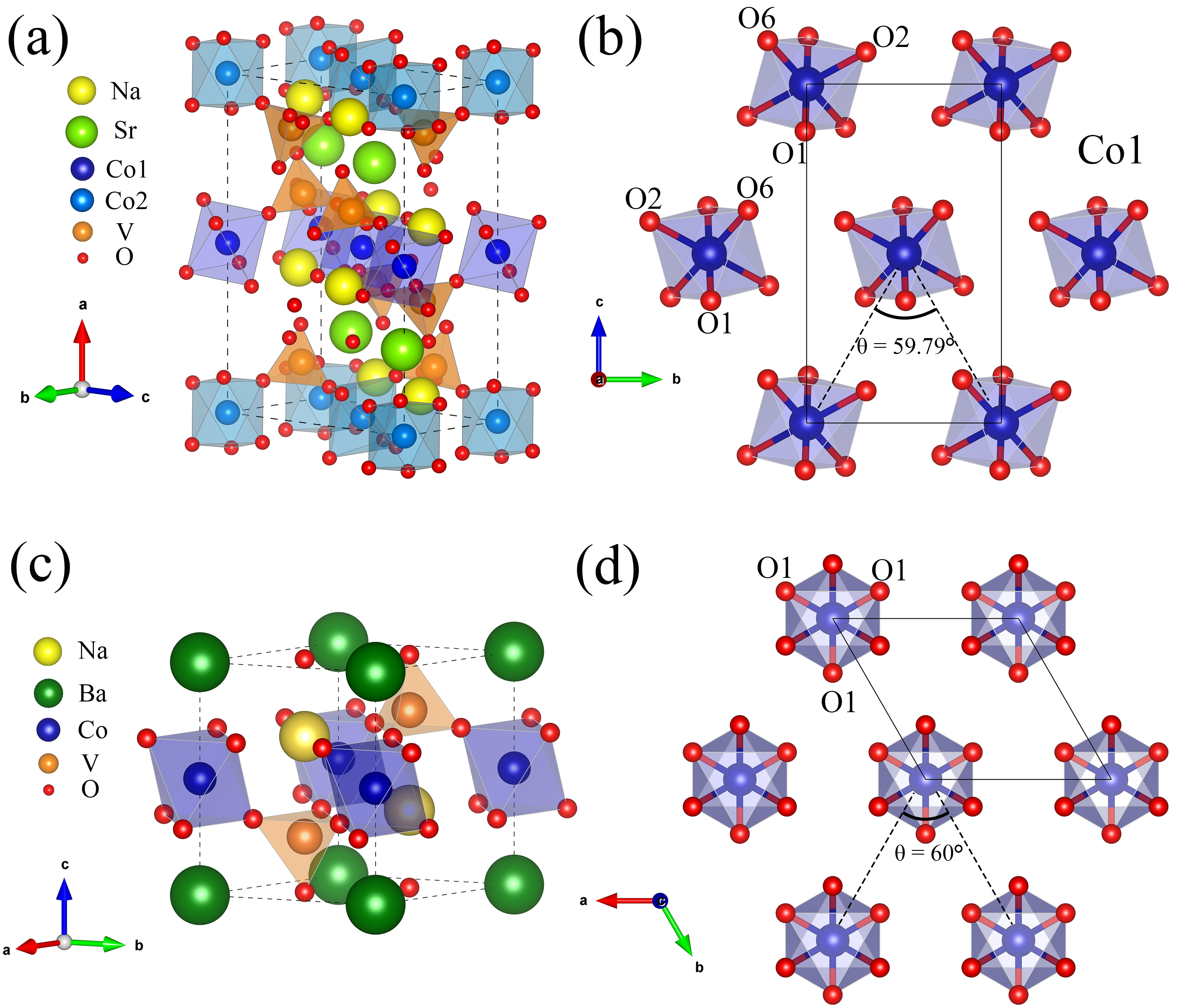}
    \caption{\label{fig.1}
    Crystal structure of NSCVO (a) with a monoclinic symmetry (space group $P \rm 2_1/c$), compared with that of NBCVO (c) with a trigonal symmetry (space group $P\bar{3}m1$)  \cite{SANJEEWA201661}. In NSCVO, $\rm Co^{2+}$ ions in the $bc$ plane form an isosceles triangular lattice with a top angle of 59.79° at room temperature. The $\rm O_3$ faces of the Co1O$_6$ octahedra defined by O1, O2 and O6 are tilted away from the $bc$ plane and also exhibit a finite in-plane rotation (b). In NBCVO, the $\rm Co^{2+}$ ions in the $ab$ plane form an equilateral triangular lattice, and $\rm O_3$ faces of the $\rm CoO_6$ octahedra, defined by three equivalent O1 atoms, are strictly parallel to the $ab$ plane (d).}.
\end{figure*}
    Despite these insights, detailed investigations of the vanadate members within this family remain limited. 
    Among them, trigonal $\rm Na_2BaCo(VO_4)_2$ (NBCVO) has been reported to exhibit ferromagnetic ordering near 4 K, supported by comprehensive experimental measurements \cite{Nakayama_2013}.
    In contrast, although the crystal structure of the Sr-analogue material, $\rm Na_2SrCo(VO_4)_2$ (NSCVO), has been reported to feature a slight monoclinic distortion \cite{SANJEEWA201661}, its magnetic properties have yet to be systematically explored.
    Substituting $\rm Ba^{2+}$ with smaller $\rm Sr^{2+}$ ion naturally modifies the local coordination environment and lattice symmetry, thereby altering the exchange pathways and anisotropy.
    Such structural and electronic variations could, in turn, stabilize a fundamentally different magnetic ground state, making NSCVO an intriguing candidate for further study. 

    In this work, we systematically investigate NSCVO, a glaserite-type cobalt vanadate in which $\rm Co^{2+}$ ions form distorted triangular layers.
    By combining x-ray diffraction, neutron powder diffraction, bulk magnetization, and heat-capacity measurements, we elucidate both the crystal structure and magnetic properties of NSCVO.
    Our results reveal that the compound undergoes a transition to long-range FM order below $T_{\rm C} \approx 3.4~ {\rm K}$.
    The  magnetization behavior and recovered magnetic entropy are consistent with an effective spin-1/2 ground state, characteristic of cobalt-based frustrated magnets.
    Our findings place NSCVO as a new member of triangular-lattice cobaltate family and provide valuable insights into the microscopic mechanisms governing magnetism in geometrically frustrated systems.
\section{METHODS}
    Polycrystalline samples of NSCVO were synthesised via a conventional solid-state reaction method.
    Stoichiometric amounts of $\rm Na_2CO_3$ ($99.99\%$), $\rm SrCO_3$ ($99.99\%$), $\rm CoO$ ($99.99\%$) and $\rm V_2O_5$ ($99.99\%$) were mixed, thoroughly ground, and pressed into pellets.
    The pellets were sintered in air atmosphere at 750 ℃ for 12 hours and furnace-cooled to room temperature.
    This sintering process, including intermediate grinding, was repeated twice to minimize possible impurities. 

    Room-temperature x-ray diffraction (XRD) and neutron powder diffraction (NPD) measurements were performed to examine the phase purity of the synthesized samples.
    XRD patterns were obtained using a Bruker D8 diffractometer with Cu $K\alpha$ radiation, $\lambda = 1.541$ \AA,
    while room-temperature NPD data were collected on the General Purpose Powder Diffractometer (31113.02.CSNS.GPPD) at China Spallation Neutron Source (CSNS, Dongguan, China) operating in a time-of-flight (TOF) mode \cite{CHEN2018370}.
    Low-temperature NPD measurements at 2.3 K and 10 K were conducted at the High Intensity Powder Diffractometer (HIPD) at China Advanced Research Reactor (CARR, Beijing, China) using the incident neutron wavelength $\lambda = 1.479$ \AA \cite{HGuo}.
    Rietveld refinements of all XRD and NPD data were performed by the \textsc{FullProf} suite \cite{RODRIGUEZCARVAJAL199355}.

    DC magnetization measurements on the polycrystalline sample were carried out using a Quantum Design Magnetic Property Measurement System MPMS3. 
    Zero-field-cooling (ZFC) magnetization versus temperature was measured from 1.8 K to 300 K under an applied field of 0.1 T.
    Isothermal magnetization curves were measured at 1.8 K in magnetic fields from $-$7 T to +7 T.
    Specific heat measurements were performed in zero field from 1.8 K to 55 K, with a Quantum Design Physical Property Measurement System PPMS.

\section{RESULTS}
\subsection{Structural characterizations}

\begin{table*}[ht]
    \caption{\label{Table I} The room-temperature structural parameters of NSCVO, including the atomic coordinates, isotropic displacement parameters ($B\rm_{iso}$), and site occupancies (Occ.), as determined by Rietveld refinements to both the XRD and NPD patterns.
    (Space group: $P2_1/c$, $a$ = 13.68434 \AA, $b$ = 5.50822 \AA, $c$ = 9.58179 \AA, $\alpha$ = 90°, $\beta$ = 89.974°, $\gamma$ = 90°)}
    \begin{ruledtabular}
    \begin{tabular}{c c c c c c c}
    Atom & Wyckoff pisition&       \emph{x}&       \emph{y}&       \emph{z}&  $B\rm_{iso}$ (\AA$^2$) &     Occ.     \\
    \hline
    Sr   &        4e       &  0.7501(5)&  0.0529(4)&  0.0387(2)&       0.90(6)     &     1.00    \\
    Na1  &        4e       &  0.9178(9)&  0.9827(11)&  0.6802(10)&       1.30(28)     &     0.95(2)    \\
    Na2  &        4e       &  0.5835(9)&  0.0179(12)&  0.3509(9)&       1.02(27)     &     0.92(2)    \\
    Co1  &        2b       &  0.5&  0&  0&       0.91(39)     &     1.00    \\
    Co2  &        2a       &  0&  0&  0&       0.10(33)     &     1.00    \\
    V1   &        4e       &  0.6386(10)&  0.964(2)&  0.6818(11)&       0     &     1.00    \\
    V2   &        4e       &  0.8630(11)&  0.985(2)&  0.3540(10)&       0     &     1.00    \\
    O1   &        4e       &  0.6172(3)&  0.9919(7)&  0.8613(5)&       0.87(9)     &     1.00    \\
    O2   &        4e       &  0.5775(4)&  0.1947(8)&  0.5949(4)&       0.89(9)     &     1.00    \\
    O3   &        4e       &  0.8814(3)&  0.6974(9)&  0.4172(4)&       1.08(10)     &     1.00    \\
    O4   &        4e       &  0.9161(3)&  0.1988(8)&  0.4584(4)&       0.88(9)     &     1.00    \\
    O5   &        4e       &  0.9253(4)&  0.9924(6)&  0.1931(5)&       1.27(10)     &     1.00    \\
    O6   &        4e       &  0.5871(3)&  0.6998(8)&  0.6280(4)&       1.08(10)     &     1.00    \\
    O7   &        4e       &  0.7535(4)&  0.9777(6)&  0.6381(5)&       1.00(10)     &     1.00    \\
    O8   &        4e       &  0.7468(4)&  0.0557(6)&  0.3204(4)&       1.15(9)     &     1.00    \\
    \end{tabular}
    \end{ruledtabular}
\end{table*}

\begin{figure}
    \centering
    \includegraphics[width = \linewidth]{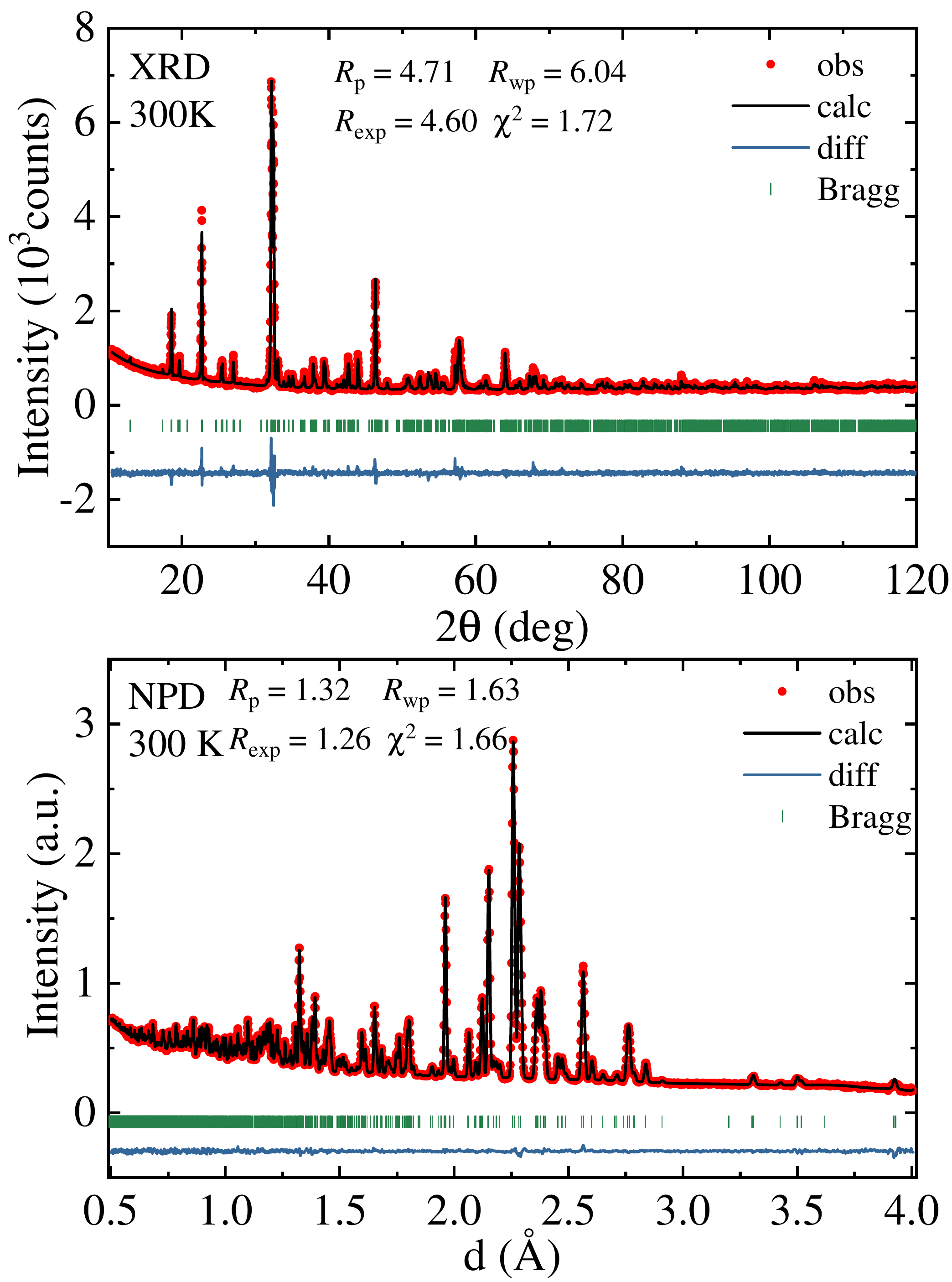}
    \caption{\label{fig.2}
    Room-temperature XRD (a) and NPD (b) patterns of polycrystalline NSCVO and the Rietveld refinements with the $R$-factors and $\chi^2$ provided. The red circles represent the observed intensities, and the black solid lines are the calculated patterns. The differences between the observed and calculated intensities are shown as blue solid lines at the bottom. The Bragg peaks of NSCVO are marked by green bars.}
\end{figure}

    The crystal structure of NSCVO was determined by combination of XRD and NPD at room temperature.
    The starting structure model was taken from Ref. \onlinecite{SANJEEWA201661}, as determined by single-crystal XRD, which is insensitive to light ions like $\rm O^{2-}$.
    The combination of XRD with NPD, which carries more information about $\rm O^{2-}$, facilitate a more percise structure determination.
    Adopting the monoclinic structure (space group $P \rm{2_1/c}$) illustrated in Fig. 1(a), both the XRD and NPD patterns can be well fitted, as shown in Figs. 2(a) and 2(b), respectively.
    The satisfactory $R$ factors and goodness of fit ($\chi^2$) indicate  the validity of the structural model.
    The refined atomic coordinates, isotropic displacement parameters ($B\rm_{iso}$) and site occupancies (Occ.) are listed in Table.1. There are slight deficiencies on both Na1 and Na2 sites, according to the refinement.
    As shown in Fig. 1(b), the $\rm Co^{2+}$ ions in the $bc$ plane form an isosceles triangle with a top angle of 59.79°. 
    Compared with the trigonal NBCVO (space group $P\bar{3}m1$) consisting of equilaterally trianglular layers (Figs.1(c) and 1(d)) \cite{SANJEEWA201661}, the substitution of Ba with Sr results in a monoclinic distortion in NSCVO due to the reduced ionic radius.
    Figs.1(b) and 1(d) highlight the contrasting local environments of the $\rm CoO_6$ octahedra in NBCVO and NSCVO.
    In NBCVO, the $\rm CoO_6$ octahedra undergo a well-defined trigonal elongation, as illustrated in Fig.1 (d) from the view along the $c$-axis.
    The two $\rm O_3$ triangular faces, defined by three equivalent O1 atoms, are strictly parallel to the $ab$ plane, consistent with the threefold rotational symmetry of the trigonal structure.
    In constrast, the distortion of the $\rm CoO_6$ octahedra in NSCVO is considerably more complex.
    Taking the $\rm Co1O_6$ octahedron as an example, Fig.1(b) shows that the $\rm O_3$ triangular faces defined by O1, O2 and O6 are tilted away from the $bc$ plane (the in-plane direction of the monoclinic $P2_1/c$ structure) and also exhibits a finite rotation within the $bc$ plane. These combined tilting and rotational components reflect the lowered symmetry and more complex local environment of the Sr compound compared with its Ba analogue, which was also observed in the case of $\rm Na_2SrCo(PO_4)_2$, where the full substitution of $\rm Sr^{2+}$ for $\rm Ba^{2+}$ also leads to a trigonal-to-monoclinic structural transformation \cite{zhang2022doping}.

\subsection{Macroscopic magnetic properties}
    The dc magnetization of polycrystalline NSCVO measured in a ZFC process from 1.8 K to 300 K under an applied magnetic field of 0.1 T agrees with a typical ferromagnetic behavior. 
    As shown in Fig.3(a), the magnetization ($M$) or static magnetic susceptibility ($\chi=M/H$) increases rapidly at low temperature below 5 K.
    Its derivative to temperature (d$\chi$/d$T$), as shown in the inset of Fig. 3(a), exhibits a pronounced dip around 3.5 K, which can be assigned as the ferromagnetic ordering temperature $T_{\rm C}$.

    Fig.3(b) shows the Curie-Weiss fitting to the inverse susceptibility (1/$\chi$) in the paramagnetic state, performed for the low-temperature range (5-10 K) and high-temperature range (200-300 K), respectively, using the formula $\chi = C/(T - \theta_{\rm CW})+\chi_0$.
    The term $\chi_0$ denotes the temperature-independent susceptibility, which includes the positive van Vleck contribution, the negative core-diamagnetic contribution, and the extrinsic diamagnetic background from the sample holder.
    For the high-temperature fitting, the estimated effective moment is $\mu_{\rm eff} =$ 5.99$~\mu\rm_B$, the Curie-Weiss temperature is $\theta_{\rm CW} =$ $-$30.77$~{\rm K}$, and $\chi_0$ = $\rm -1.5\times 10^{-3} ~emu~Oe^{-1}~mol^{-1}$.
    This value of $\mu_{\rm eff}$ is significantly larger than the theoretical value of $3.87~ \mu\rm_B$ for high-spin $\rm Co^{2+}$ ions.
    Such enhancement arises from the substantial orbital contributions and, more importantly, from the thermal population of the SOC-split Kramers doublets at high temperatures, which have been discussed in other similar Co-based materials \cite{PhysRevLett.125.047201}.
    On the other hand, the low-temperature fitting yields $\mu_{\rm eff} = 4.43~\mu\rm_B$, $\theta_{\rm CW} =$ 3.27$~{\rm K}$, if adopting the same value of $\chi_0$.
    In this low-temperature regime, only the lowest Kramers doublet is populated.
    In analogy to NBCPO, where INS measurements revealed a first excited doublet around 40 meV \cite{PhysRevLett.134.136703} , it is reasonable to assume a similar crystal-field/SOC energy scale in NSCVO. Therefore, whereas the low-temperature behavior is governed by an effective $S\rm_{eff}$ = 1/2 ground state, as commonly observed in Co-based compounds \cite{PhysRevLett.110.267201,PhysRevLett.125.047201,doi:10.1073/pnas.1906483116,xu2024frustrated,Nakayama_2013}. The reduction of $\mu_{\rm eff}$ upon cooling is thus a consequence of the depopulation of the excited SOC levels rather than a change in the intrinsic spin state.
    More importantly, the positive low-temperature Curie-Weiss temperature of $\theta_{\rm CW}$ = 3.27$~{\rm K}$ indicates dominant ferromagnetic interactions between the $\rm Co^{2+}$ ions in NSCVO, consistent with the intralayer ferromagnetic order as confirmed by NPD below.
\begin{figure}[ht]
    \centering
    \includegraphics[width = 0.9\linewidth]{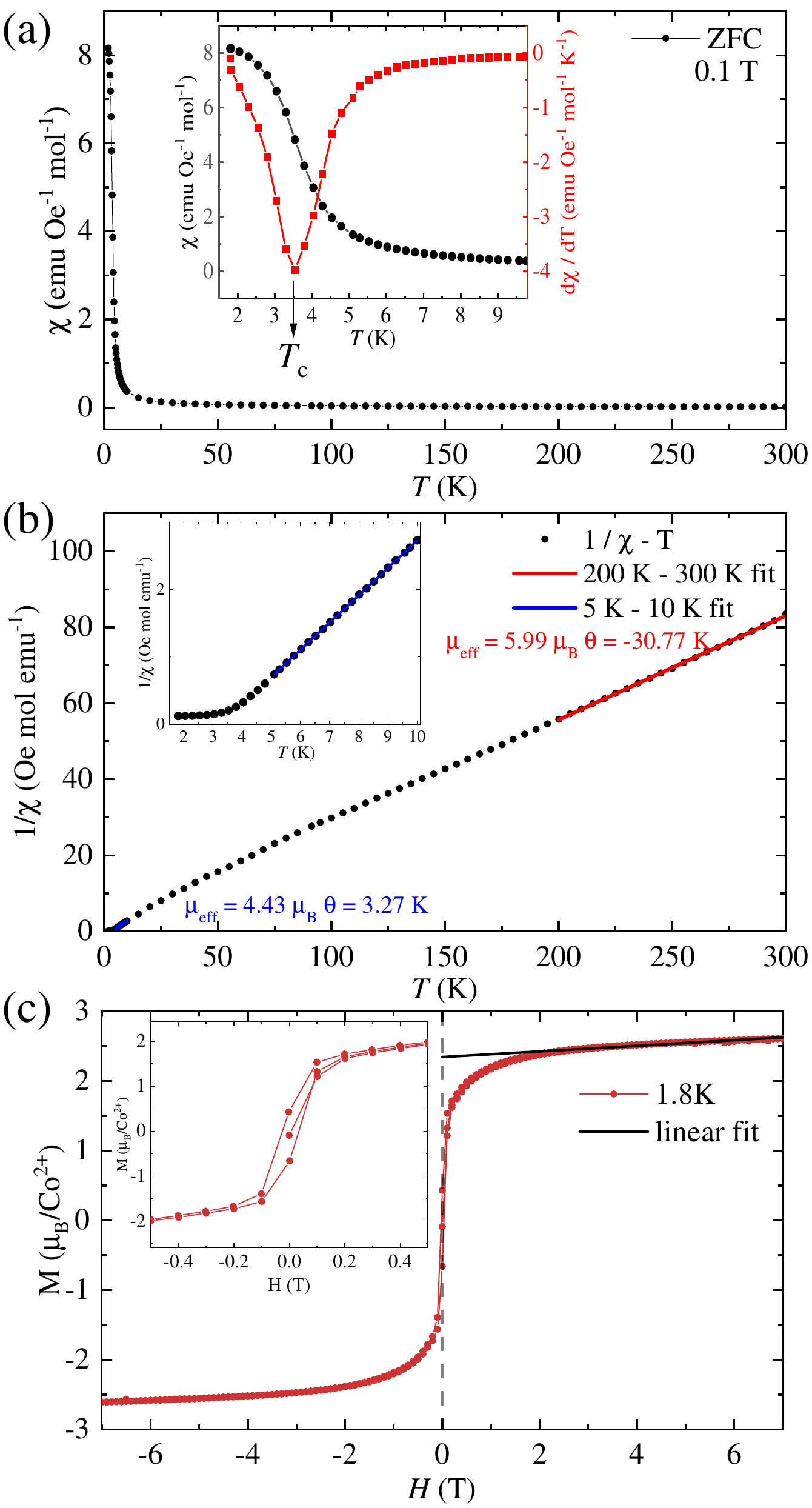}
    \caption{\label{fig.3}
    (a) DC magnetic susceptibility ($\chi$) of polycrystalline NSCVO, measured in a magnetic field of 0.1 T.
    The inset shows the low-temperature part of $\chi$ below 10 K and its first derivative (d$\chi$/d$T$) with a dip around 3.5 K.
    (b) The inverse magnetic susceptibility (1/$\chi$) and Curie-Weiss fittings to the high-temperature (200-300 K) and low-temperature (5-10 K) ranges, as represented by the red and solid lines, respectively.  The inset shows an enlarged view of the low-temperature region.
    (c) Isothermal magnetization of NSCVO measured at 1.8 K, with the low-field hysteresis loop highlighted in the inset. The solid line represents a linear extrapolation to subtract the Van Vleck paramagnetism.} 
\end{figure}

    Fig.3(c) shows the isothermal magnetization curve of NSCVO measured at 1.8 K, where a rapid increase below 0.1 T is observed and a saturation of magnetization is reached above $\rm \sim 1.5~T$.
    After subtracting the sloping background due to the Van Vleck paramagnetism, the saturation magnetic moment is estimated to be $\rm 2.34~\mu_B/Co^{2+}$.
    In addition, a hysteresis loop observed in the low-field region, as shown in the inset of Fig.3(c), further confirms its ferromagnetic ground state.

\subsection{Specific heat}

\begin{figure}
    \centering
    \includegraphics[width = \linewidth]{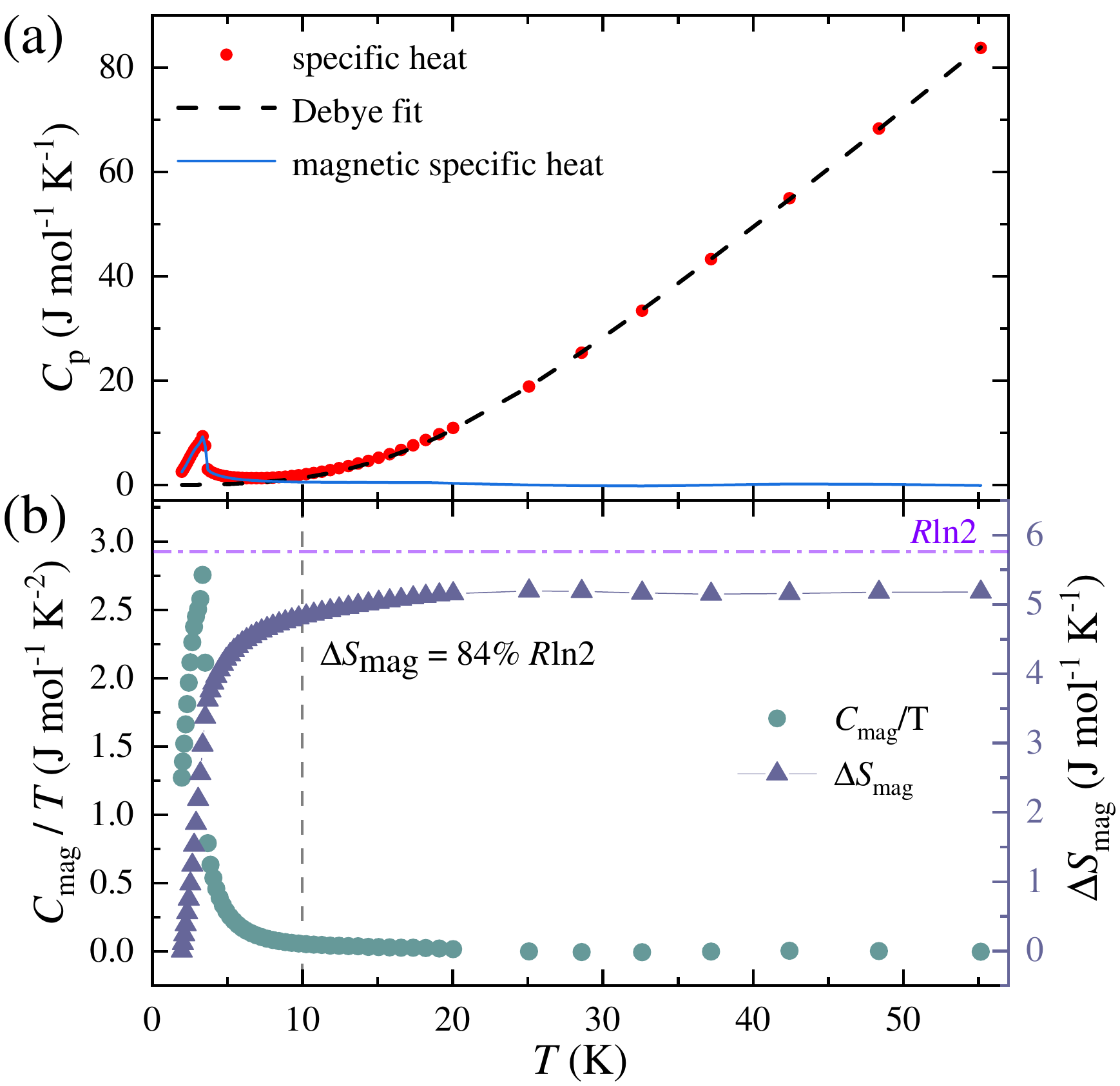}
    \caption{\label{fig.4}
    (a) Specific heat ($C_{\rm p}$, red circles) of polycrystalline NSCVO. The black dashed line is an approximation to the phonon specific heat ($C\rm_{ph}$) using a two-component Debye model, and the blue solid line corresponds to the magnetic specific heat ($C\rm_{mag}$) after substracting the phonon contribution.
    (b) $C_{\rm mag}/T$ (green circles) and the estimated magnetic entropy change ($\Delta S_{\rm mag}$, purple triangles) as functions of temperature, respectively. The horizontal dashed line marks $R{\rm ln}2$ as expected for a spin-1/2 system. The vertical dashed line marks 10 K, where most of the entropy is already recovered and $\Delta S_{\rm mag}$ reaches $0.84R{\rm ln2}$.}
\end{figure}

    Fig.4 presents the temperature dependence of the specific heat, $C_{\rm p}(T)$, of the polycrystalline NSCVO sample measured in zero magnetic field.
    A $\lambda$-shaped peak at $T$ = 3.4(1)~K signals the occurrence of a long-range magnetic order of $\rm Co^{2+}$, consistent with the result from the magnetization measurements.
    The phonon contribution to the specific heat is approximated using a two-component Debye model assuming two different Debye temperatures, as widely adopted in many other materials \cite{somesh2021,ranjith2017,Su2023}, by dividing the atoms into a lighter and a heavier group, respectively, 
    $$ C\rm_{ph} = 9\it{R}\sum_{i=\rm1}^{\rm2} {c_i (\frac{T}{\theta_{Di}})^{\rm3} \int_{\rm0}^{\theta_{Di}/T}{ \mathrm{d}x  \frac{x^{\rm4} e^x}{(e^x-\rm1)^{2\rm}} } ,}$$
    where $c_i$ are weighting factors (number of atoms associated with each Debye component) and $\theta_{Di}$ denote the two different Debye temperatures.
    Since there are 14 atoms in total per chemical formula unit, the constraint $c_1 + c_2 = 14$ was imposed in the fitting.
    The best fit parameters are found to be $c_1 = 3.08$, $c_2 = 10.92$, $\theta_{\rm D1} = 172.58$ K, $\theta_{\rm D2} = 456.92$ K.
    The values of $c_1$ and $c_2$ roughly match the numbers of heavier atoms (Sr and Co) and lighter atoms (Na, O and V).

    The magnetic specific heat ($C_{\rm mag}$) is then obtained, as presented by the solid line in Fig. 4(a), by subtracting the estimated phonon specific heat $C_{\rm_{ph}}$ (the dashed line in Fig.4(a)), and the magnetic entropy change ($\Delta S_{\rm mag}$) is calculated by intergrating $C_{\rm mag}/T$ over the temperature.
    As Fig.4(b) shows, the experimental maximal change of magnetic entropy is estimated to be ~5.18 J K$^{-1}$ mol$^{-1}$ up to 55 K, which is about $89.93\%$ of $R{\rm ln2}$. Importantly, most of the entropy is already recovered at much lower temperatures: by 10 K, $\Delta S_{\rm mag}$
 reaches 4.84 J K$^{-1}$ mol$^{-1}$, which corresponds to $84\%$ of $R{\rm ln2}$.
    This indicates that the essential magnetic degrees of freedom associated with the lowest Kramers doublet of $\rm Co^{2+}$ ions are largely released near the ordering temperature, consistent with an effective spin-1/2 description commonly found in cobalt-based frustrated magnets \cite{zhang2022doping,xiang2024giant,kojima2018quantum,lin2021field,xu2024frustrated}.
    The remaining $\sim 10\%$ shortfall may arise from an overestimation of the phonon specific heat based on our two-component Debye model or missing entropy residing in higher crystal-field levels, rather than strong magnetic fluctuations above the transition. Overall, the entropy evolution is fully consistent with a well-isolated Kramers doublet governing the low-temperature magnetism.

\subsection{Magnetic structure determination}

\begin{figure}
    \centering
    \includegraphics[width = \linewidth]{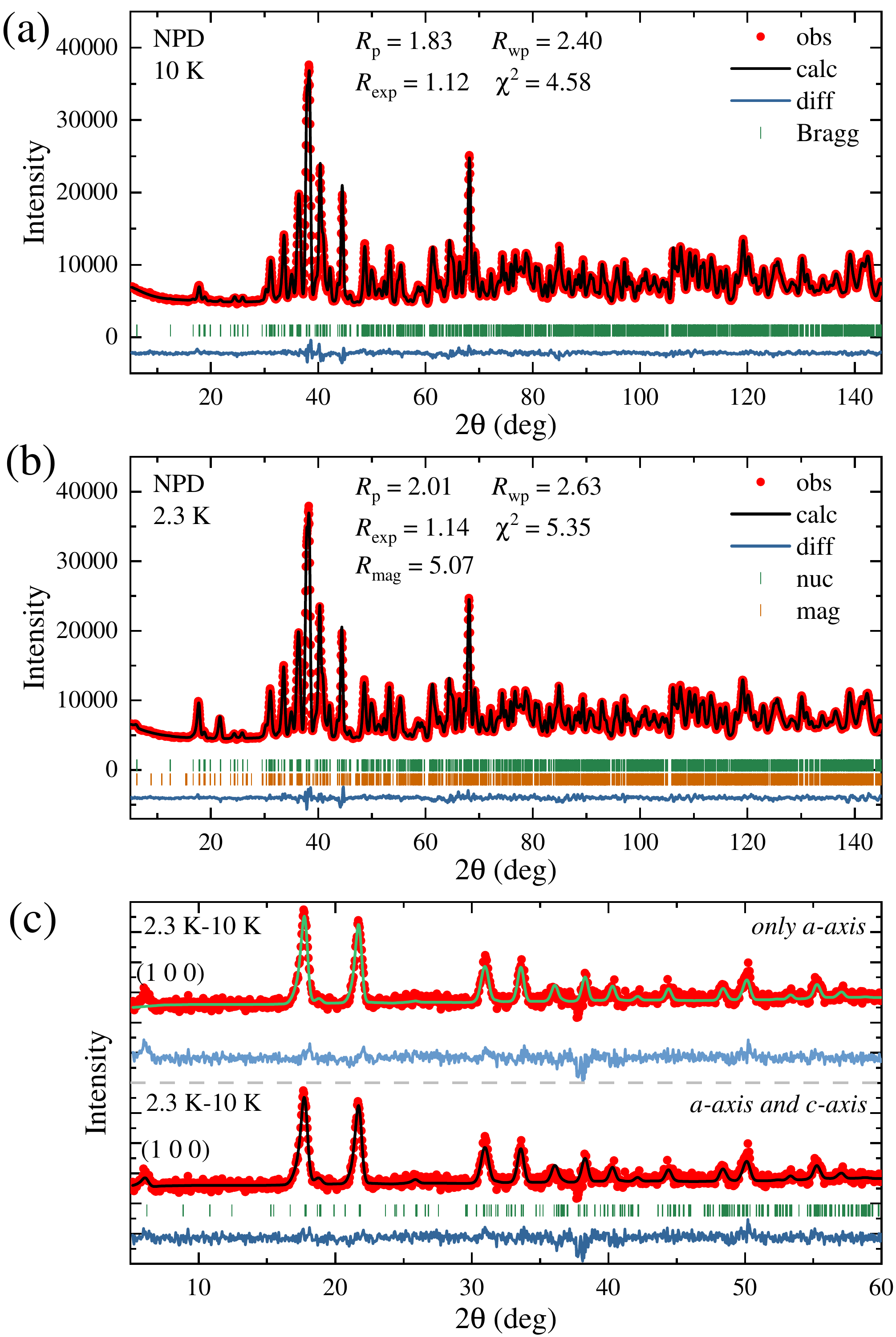}
    \caption{\label{fig.5}
    Low-temperature NPD patterns collected at 10 K (a) and 2.3 K (b) as well as the Rietveld refinements. The nuclear and magnetic reflections are marked by green and brown vertical bars, respectively. After subtracting the nuclear scattering (10 K) from the 2.3 K data set, the net magnetic scattering (2.3 K$-$10 K) is obtained and fitted using different magnetic structures described by $\Gamma_1$, with the $a$-axis component only (top) and both $a$-axis and $c$-axis components (bottom), respectively. Clearly, the latter well accounts for the intensity of (1 0 0) magnetic reflection.} 
\end{figure}

\begin{figure}[b]
    \centering
    \includegraphics[width = \linewidth]{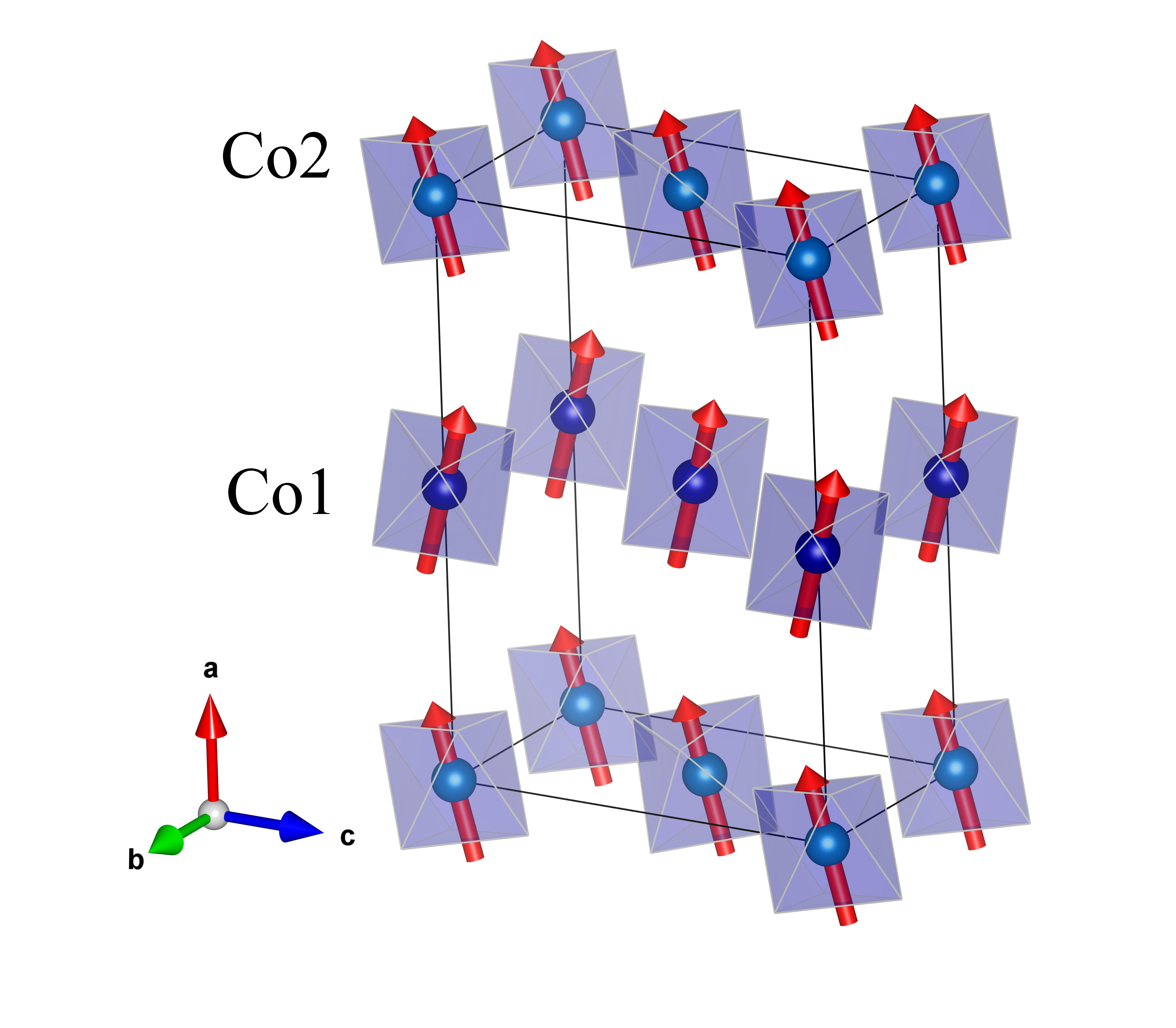}
    \caption{\label{fig.6} 
    Canted FM structure of NSCVO determined by NPD at 2.3 K, wtih
    $\rm Co1$ and $\rm Co2$ denoted by different colors.}
\end{figure}

\begin{table}[t]
    \caption{\label{Table III} Basis vectors of the irreducible representations (IRs) of different Co sites in NSCVO with the propagation vector $k = (0, 0, 0)$.
    }
    \begin{ruledtabular}
    \begin{tabular}{c c c c c c}
              &          &                             \multicolumn{4}{c}{Basis Vectors}                         \\\cline{3-6}
              &          &        \multicolumn{2}{c}{Site(2b)}       &        \multicolumn{2}{c}{Site(2a)}       \\\cline{3-6} 
              &          & Co1\textunderscore1 & Co1\textunderscore2 & Co2\textunderscore1 & Co2\textunderscore2 \\
              &          &       (0.5 0 0)     &     (0.5 0.5 0.5)   &       (0 0 0)       & (0 0.5 0.5)         \\\hline
              & $\Psi_1$ &       (1 0 0)       &        (1 0 0)      &       (1 0 0)       & (1 0 0)             \\
    $\Gamma_1$ & $\Psi_2$ &       (0 1 0)      &        (0 $-$1 0)     &       (0 1 0)       & (0 $-$1 0)            \\
              & $\Psi_3$ &       (0 0 1)       &        (0 0 1)      &       (0 0 1)       & (0 0 1)             \\\hline
              & $\Psi_1$ &       (1 0 0)       &        ($-$1 0 0)     &       (1 0 0)       & ($-$1 0 0)            \\
    $\Gamma_2$ & $\Psi_2$ &       (0 1 0)      &        (0 1 0)      &       (0 1 0)       & (0 1 0)             \\
              & $\Psi_3$ &       (0 0 1)       &        (0 0 $-$1)     &       (0 0 1)       & (0 0 $-$1)            \\
    \end{tabular}
    \end{ruledtabular}
\end{table}

    To determine the magnetic structure of NSCVO, the NPD data were collected at 10 K (well above $T_{\rm C} \approx 3.4~{\rm K}$) and 2.3 K (below $T_{\rm C}$), respectively.
    Fig.5(a) shows the NPD pattern in the paramagnetic state at 10 K .
    It can be well fitted by the same crystallographic model that was established for room temperature, indicating no detectable structural phase transition within the experimental resolution.
    Upon cooling to the magnetically ordered state at 2.3 K, additional magnetic scatterings emerge (see Fig.5(b)), which are better visualized in the difference between 2.3 K and 10 K, as shown in Fig.5(c).
    As a result, the magnetic reflections coincide with the nuclear Bragg peaks and can therefore be indexed with a magnetic propagation vector $k = (0, 0, 0)$, implying that the magnetic unit cell is as large as the crystallographic cell. The widths of magnetic reflections are comparable to the nuclear Bragg peaks, indicating fully three-dimensional long-range magnetic correlations.
\begin{table*}
    \caption{\label{Table IV} Summary of key structural and magnetic properties of Co-based compounds in the family of $X_2Y$Co($T$O$_4)_2$ ($X$ = Na, Ag; $Y$ = Sr, Ba; $T$ = V, P).}
    \begin{ruledtabular}
    \begin{tabular}{c c c c c c}
    materials                  &  space group (300 K)  &     type of interactions    &       $T_{\rm N}$/$T_{\rm C}$ (K)     &          reference         \\
    $\rm Na_2SrCo(VO_4)_2$     &     $P2_{1}/c$        &      FM          &          3.4            &       This work      \\
    $\rm Na_2BaCo(VO_4)_2$     &     $P\bar{3}$        &      FM          &          4       &       \cite{doi:10.1021/acs.inorgchem.8b03418} \\
    $\rm Ag_2SrCo(VO_4)_2$     &       $C2/c$          &      FM          &          3.8     &       \cite{PhysRevB.85.214422}        \\
    $\rm Ag_2BaCo(VO_4)_2$     &      $P\bar{3}$       &      FM          &          4.2     &       \cite{PhysRevB.85.214422}          \\
    $\rm Na_2SrCo(PO_4)_2$     &     $P2_{1}/c$        &      AFM         &          $\sim$ 0.6     &       \cite{PhysRevB.106.054415}          \\
    $\rm Na_2BaCo(PO_4)_2$     &      $P\bar{3}m1$ or $P\bar{3}$  &      AFM         &          $\sim$ 0.15    &       \cite{xiang2024giant,1pvl-kzjm,Kajita2024}           \\
    \end{tabular}
    \end{ruledtabular}
\end{table*}    
    The program \textsc{BasIreps} integrated into the \textsc{FullProf} suite was employed to deduce the symmetry-allowed irreducible representations (IRs).
    In the unit cell, there are two crystallographically distinct $\rm Co^{2+}$ sites: Co1 locating at (0.5, 0, 0) and symmetry-equivalent (0.5, 0.5, 0.5), and Co2 residing at (0, 0, 0) and (0, 0.5, 0.5).
    For the space group $P2_1/c$ and $k = (0, 0, 0)$, symmetry analysis by \textsc{BasIreps} yields two candidate IRs for the magnetic moments on the Co sites, whose basis vectors are listed in Table II. 
    For $\Gamma_1$, the Co$1\textunderscore1$ and Co$1\textunderscore2$ moments, locating at (0.5 0 0) and (0.5 0.5 0.5), respectively, are aligned ferromagnetically along $a$ and $c$ axes, but antiferromagnetically along $b$ axis (for Co$2\textunderscore1$ and Co$2\textunderscore2$, the coupling adopts the same way).
    In contrast, for $\Gamma_2$, the coupling between them is AFM along $a$ and $c$ axes but FM along $b$ axis.

    In our case, the magnetic structure described by $\Gamma_1$ yields a much better agreement with the experimental data, compared with the model given by $\Gamma_2$.
    The corresponding fitting to the observed magnetic diffraction intensities using $\Gamma_1$ is shown in Fig.5(c).
    Obviously, both the $a$-axis and $c$-axis components of the Co$^{2+}$ moments have to be non-zero, as adopting a similar FM structure of NBCVO with only the out-of-plane component \cite{doi:10.1021/acs.inorgchem.8b03418} fails to reproduce the lowest-angle (1 0 0) magnetic reflection. The $b$-axis component converges to almost zero in the refinement.
    The corresponding determined magnetic structure is depicted in Fig.6. 
    The Co$^{2+}$ moments are canted and ferromagnetically aligned within the triangular-lattice layers, while adjacent layers are coupled antiferromagnetically along the $c$ axis.
    The refined $a$-axis and $c$-axis components of the ordered moments are $m\rm_{a}(Co1) = 2.554~\mu_B$, $m\rm_{c}(Co1) = 0.572~\mu_B$ and $m\rm_{a}(Co2) = 2.530~\mu_B$, $m\rm_{c}(Co2) = 0.687~\mu_B$, yielding total magnetic moments of $m\rm(Co1) = 2.617~\mu_B$ and $m\rm(Co2) = 2.622~\mu_B$, respectively.
    It is worth pointing out that our lowest temperature of 2.3 K might not be low enough compared with $T_{\rm C} \approx 3.4~{\rm K}$ and the fully saturation magnetic moment in the real ground state may be even larger.
    In addition, according to our refinement, the Co$^{2+}$ moments are basically staggered aligned in the $ac$ plane and no $b$-axis components can be identified within the experimental resolution.
    Such a magnetic structure well accounts for the net ferromagnetism observed in the magnetization measurements.

    Interestingly, the refined moment direction is broadly consistent with the tilting of the CoO$_6$ octahedra, whose O$_3$ triangular face is tilted by about 10° away from the $bc$-plane, although the alignment is not perfect. This slight deviation is expected, as the monoclinic distortion introduces additional crystal-field and exchange-anisotropy contributions beyond a simple trigonal scenario.


%

\section{DISCUSSIONS AND SUMMARY}

    The canted ferromagnetic structure of monoclinic NSCVO with a slight tilting of the Co$^{2+}$ moments off its $a$ axis perpendicular to the triangular-lattice layers is in contrast to that of trigonal NBCVO, in which the Co$^{2+}$ moments align ferromagnetically purely along the out-of-plane direction.
    As a comparison, the trigonal NBCPO shows a Y-like spin supersolid antiferromagnetic ground state, while the monoclinic NSCPO was also reported to order antiferromagnetically with its detailed spin structure unreported yet \cite{xiang2024giant,PhysRevB.106.054415}.
    A wide variety of spin configurations in these glaserite-type vanadates and phosphates indicates the important role of the detailed crystallographic environment of the Co$^{2+}$ ions, especially the VO$_4$ or PO$_4$ tetrahedra, in mediating their exchange interactions and corresponding spin configurations.

    The Curie temperature ($T_{\rm C} \approx 3.4~ {\rm K}$) of NSCVO is slightly lower, compared with that of its sister compound NBCVO with $T_{\rm c} \approx 4~ {\rm K}$ \cite{Nakayama_2013}. 
    Considering the slight monoclinic distortion induced by the Ba/Sr substitution, the decrease of $T_c$ in NSCVO is reminiscent of similar trends those observed in other vanadates $\rm Ag_2SrCo(VO_4)_2$/ $\rm Ag_2BaCo(VO_4)_2$ and $\rm Ag_2SrNi(VO_4)_2$/ $\rm Ag_2BaNi(VO_4)_2$ \cite{PhysRevB.85.214422}, where the substitution of the smaller AE$^{2+}$ cation leads to a reduction in the lattice symmetry and a decrease in $T_{\rm c}$, as listed in Table.IV. 
    In contrast, for the phosphastes NSCPO and NBCPO, the monoclinic distortion of the triangular lattice due to the Ba/Sr substitution, however, increase $T_{\rm N}$ by a factor of four \cite{xiang2024giant,zhang2022doping,PhysRevB.106.054415}. 

    In the glaserite structure, the CoO$_6$ octahedra neither share edges nor corners, and the Co-Co distances exceed 5 Å. Therefore, the direct Co-O-O-Co super-superexchange is expected to be weak. The dominant exchange pathway is generally considered to be Co-O-$T$-O-Co ($T$=V, P), in which the $T$O$_4$ tetrahedra mediate the superexchange.
    The distinct behaviors of vanadates and phosphates can be attributed to the different roles of the VO$_4$ and PO$_4$ tetrahedra in adjusting the superexchange pathways between $\rm Co^{2+}$ ions. 
    NBCVO and NBCPO serve as good examples to illustrate this behavior.
    Although the intralayer and interlayer Co-Co distances in NBCVO (5.5489 Å and 7.0741 Å) are significant larger than those in NBCPO (5.3134 Å and 7.0076 Å), the Co-X-Co bond angles are nearly identical (96.84° vs 97.30°), which should favor ferromagnetic interactions according to the Goodenough-Kanamori-Anderson (GKA) rules.
    The key difference therefore lies in the electronic structure of the nonmagnetic tetrahedra. $\rm VO_4^{3-}$ contains a $\rm V^{5+}$  ($3d^{\rm 0}$) center whose empty and spatially extended $3d$ orbitals strongly hybridize with $\rm O^{2-}$ $2p$ orbitals, providing an efficient virtual-hopping channel that enhances the ferromagnetic component of the Co-O-V-O-Co superexchange.
    In contrast, the empty $3d$ orbitals of $\rm P^{5+}$ ion in $\rm PO_4^{3-}$ are much more deeply bound and spatially contracted, leading to very weak P $3p$–O $2p$ hybridization and suppressing the corresponding ferromagnetic exchange pathway. This fundamental difference in orbital hybridization naturally results in the contrasting magnetic ground states of the vanadates (FM) and phosphates (AFM).

    In antiferromagnetic phosphates, monoclinic distortion induced by Ba/Sr substitution release the geometrical frustration to some extent, which naturally lifts the Neel temperature $T_{\rm N}$.
    However, triangular-lattice ferromagnetic vanadates do not suffer from significant frustration effect, and the monoclinic distortion may weaken the dominant ferromagnetic superexchange strength by stretching some bond lengths, which lowers the Curie temperature $T_{\rm C}$.  

    In summary, we have conducted an in-depth study of the structure and magnetism of the glaserite-type triangular-lattice cobalte vanadate $\rm Na_2SrCo(VO_4)_2$, 
    combining x-ray diffraction, magnetization, specific-heat, and neutron powder diffraction measurements.
    Compared with its barium-containing analog $\rm Na_2BaCo(VO_4)_2$, the full substitution of Ba with Sr lowers the crystal symmetry from trigonal to monoclinic and stabilizes a canted ferromagnetic order below $T_{\rm C} \approx {\rm 3.4~K}$. 
    Both magnetization and entropy analysis confirm the realization of an effective spin-1/2 state at low temperature. 
    Our results highlight the crucial role of the local environment associated with the CoO$_6$ octahedra in mediating their exchange interactions and corresponding spin structures.
    
\begin{acknowledgments}
    We acknowledge the financial supports from the National Key Projects for Research and Development of China (Grant No. 2023YFA1406003), the National Natural Science Foundation of China (Grant No. 12574154, 12074023, 12404180),  the Large Scientific Facility Open Subject of Songshan Lake (Dongguan, Guangdong, No. KFKT2022B05), and the Fundamental Research Funds for the Central Universities in China.
    This work is also supported by the Synergetic Extreme Condition User Facility (SECUF, https://cstr.cn/31123.02.SECUF).

\end{acknowledgments}




\bibliography{reference_new}

\begin{thebibliography}{52}%
\makeatletter
\providecommand \@ifxundefined [1]{%
 \@ifx{#1\undefined}
}%
\providecommand \@ifnum [1]{%
 \ifnum #1\expandafter \@firstoftwo
 \else \expandafter \@secondoftwo
 \fi
}%
\providecommand \@ifx [1]{%
 \ifx #1\expandafter \@firstoftwo
 \else \expandafter \@secondoftwo
 \fi
}%
\providecommand \natexlab [1]{#1}%
\providecommand \enquote  [1]{``#1''}%
\providecommand \bibnamefont  [1]{#1}%
\providecommand \bibfnamefont [1]{#1}%
\providecommand \citenamefont [1]{#1}%
\providecommand \href@noop [0]{\@secondoftwo}%
\providecommand \href [0]{\begingroup \@sanitize@url \@href}%
\providecommand \@href[1]{\@@startlink{#1}\@@href}%
\providecommand \@@href[1]{\endgroup#1\@@endlink}%
\providecommand \@sanitize@url [0]{\catcode `\\12\catcode `\$12\catcode
  `\&12\catcode `\#12\catcode `\^12\catcode `\_12\catcode `\%12\relax}%
\providecommand \@@startlink[1]{}%
\providecommand \@@endlink[0]{}%
\providecommand \url  [0]{\begingroup\@sanitize@url \@url }%
\providecommand \@url [1]{\endgroup\@href {#1}{\urlprefix }}%
\providecommand \urlprefix  [0]{URL }%
\providecommand \Eprint [0]{\href }%
\providecommand \doibase [0]{https://doi.org/}%
\providecommand \selectlanguage [0]{\@gobble}%
\providecommand \bibinfo  [0]{\@secondoftwo}%
\providecommand \bibfield  [0]{\@secondoftwo}%
\providecommand \translation [1]{[#1]}%
\providecommand \BibitemOpen [0]{}%
\providecommand \bibitemStop [0]{}%
\providecommand \bibitemNoStop [0]{.\EOS\space}%
\providecommand \EOS [0]{\spacefactor3000\relax}%
\providecommand \BibitemShut  [1]{\csname bibitem#1\endcsname}%
\let\auto@bib@innerbib\@empty
\bibitem [{\citenamefont {Starykh}(2015)}]{Starykh_2015}%
  \BibitemOpen
  \bibfield  {author} {\bibinfo {author} {\bibfnamefont {O.~A.}\ \bibnamefont
  {Starykh}},\ }\bibfield  {title} {\bibinfo {title} {Unusual ordered phases of
  highly frustrated magnets: a review},\ }\href
  {https://doi.org/10.1088/0034-4885/78/5/052502} {\bibfield  {journal}
  {\bibinfo  {journal} {Rep. Prog. Phys.}\ }\textbf {\bibinfo {volume} {78}},\
  \bibinfo {pages} {052502} (\bibinfo {year} {2015})}\BibitemShut {NoStop}%
\bibitem [{\citenamefont {Wosnitza}\ \emph {et~al.}(2016)\citenamefont
  {Wosnitza}, \citenamefont {Zvyagin},\ and\ \citenamefont
  {Zherlitsyn}}]{Wosnitza_2016}%
  \BibitemOpen
  \bibfield  {author} {\bibinfo {author} {\bibfnamefont {J.}~\bibnamefont
  {Wosnitza}}, \bibinfo {author} {\bibfnamefont {S.~A.}\ \bibnamefont
  {Zvyagin}},\ and\ \bibinfo {author} {\bibfnamefont {S.}~\bibnamefont
  {Zherlitsyn}},\ }\bibfield  {title} {\bibinfo {title} {Frustrated magnets in
  high magnetic fields—selected examples},\ }\href
  {https://doi.org/10.1088/0034-4885/79/7/074504} {\bibfield  {journal}
  {\bibinfo  {journal} {Rep. Prog. Phys.}\ }\textbf {\bibinfo {volume} {79}},\
  \bibinfo {pages} {074504} (\bibinfo {year} {2016})}\BibitemShut {NoStop}%
\bibitem [{\citenamefont {Khatua}\ \emph {et~al.}(2023)\citenamefont {Khatua},
  \citenamefont {Sana}, \citenamefont {Zorko}, \citenamefont {Gomilšek},
  \citenamefont {Sethupathi}, \citenamefont {Rao}, \citenamefont {Baenitz},
  \citenamefont {Schmidt},\ and\ \citenamefont {Khuntia}}]{KHATUA20231}%
  \BibitemOpen
  \bibfield  {author} {\bibinfo {author} {\bibfnamefont {J.}~\bibnamefont
  {Khatua}}, \bibinfo {author} {\bibfnamefont {B.}~\bibnamefont {Sana}},
  \bibinfo {author} {\bibfnamefont {A.}~\bibnamefont {Zorko}}, \bibinfo
  {author} {\bibfnamefont {M.}~\bibnamefont {Gomilšek}}, \bibinfo {author}
  {\bibfnamefont {K.}~\bibnamefont {Sethupathi}}, \bibinfo {author}
  {\bibfnamefont {M.~R.}\ \bibnamefont {Rao}}, \bibinfo {author} {\bibfnamefont
  {M.}~\bibnamefont {Baenitz}}, \bibinfo {author} {\bibfnamefont
  {B.}~\bibnamefont {Schmidt}},\ and\ \bibinfo {author} {\bibfnamefont
  {P.}~\bibnamefont {Khuntia}},\ }\bibfield  {title} {\bibinfo {title}
  {Experimental signatures of quantum and topological states in frustrated
  magnetism},\ }\href
  {https://doi.org/https://doi.org/10.1016/j.physrep.2023.09.008} {\bibfield
  {journal} {\bibinfo  {journal} {Phys. Rep.}\ }\textbf {\bibinfo {volume}
  {1041}},\ \bibinfo {pages} {1} (\bibinfo {year} {2023})}\BibitemShut
  {NoStop}%
\bibitem [{\citenamefont {Balents}(2010)}]{balents2010spin}%
  \BibitemOpen
  \bibfield  {author} {\bibinfo {author} {\bibfnamefont {L.}~\bibnamefont
  {Balents}},\ }\bibfield  {title} {\bibinfo {title} {Spin liquids in
  frustrated magnets},\ }\href@noop {} {\bibfield  {journal} {\bibinfo
  {journal} {Nature}\ }\textbf {\bibinfo {volume} {464}},\ \bibinfo {pages}
  {199} (\bibinfo {year} {2010})}\BibitemShut {NoStop}%
\bibitem [{\citenamefont {Zhou}\ \emph {et~al.}(2017)\citenamefont {Zhou},
  \citenamefont {Kanoda},\ and\ \citenamefont {Ng}}]{RevModPhys.89.025003}%
  \BibitemOpen
  \bibfield  {author} {\bibinfo {author} {\bibfnamefont {Y.}~\bibnamefont
  {Zhou}}, \bibinfo {author} {\bibfnamefont {K.}~\bibnamefont {Kanoda}},\ and\
  \bibinfo {author} {\bibfnamefont {T.-K.}\ \bibnamefont {Ng}},\ }\bibfield
  {title} {\bibinfo {title} {Quantum spin liquid states},\ }\href
  {https://doi.org/10.1103/RevModPhys.89.025003} {\bibfield  {journal}
  {\bibinfo  {journal} {Rev. Mod. Phys.}\ }\textbf {\bibinfo {volume} {89}},\
  \bibinfo {pages} {025003} (\bibinfo {year} {2017})}\BibitemShut {NoStop}%
\bibitem [{\citenamefont {Broholm}\ \emph {et~al.}(2020)\citenamefont
  {Broholm}, \citenamefont {Cava}, \citenamefont {Kivelson}, \citenamefont
  {Nocera}, \citenamefont {Norman},\ and\ \citenamefont
  {Senthil}}]{broholm2020quantum}%
  \BibitemOpen
  \bibfield  {author} {\bibinfo {author} {\bibfnamefont {C.}~\bibnamefont
  {Broholm}}, \bibinfo {author} {\bibfnamefont {R.~J.}\ \bibnamefont {Cava}},
  \bibinfo {author} {\bibfnamefont {S.}~\bibnamefont {Kivelson}}, \bibinfo
  {author} {\bibfnamefont {D.}~\bibnamefont {Nocera}}, \bibinfo {author}
  {\bibfnamefont {M.}~\bibnamefont {Norman}},\ and\ \bibinfo {author}
  {\bibfnamefont {T.}~\bibnamefont {Senthil}},\ }\bibfield  {title} {\bibinfo
  {title} {Quantum spin liquids},\ }\href@noop {} {\bibfield  {journal}
  {\bibinfo  {journal} {Science}\ }\textbf {\bibinfo {volume} {367}},\ \bibinfo
  {pages} {eaay0668} (\bibinfo {year} {2020})}\BibitemShut {NoStop}%
\bibitem [{\citenamefont {Wen}\ \emph {et~al.}(2019)\citenamefont {Wen},
  \citenamefont {Yu}, \citenamefont {Li}, \citenamefont {Yu},\ and\
  \citenamefont {Li}}]{wen2019experimental}%
  \BibitemOpen
  \bibfield  {author} {\bibinfo {author} {\bibfnamefont {J.}~\bibnamefont
  {Wen}}, \bibinfo {author} {\bibfnamefont {S.-L.}\ \bibnamefont {Yu}},
  \bibinfo {author} {\bibfnamefont {S.}~\bibnamefont {Li}}, \bibinfo {author}
  {\bibfnamefont {W.}~\bibnamefont {Yu}},\ and\ \bibinfo {author}
  {\bibfnamefont {J.-X.}\ \bibnamefont {Li}},\ }\bibfield  {title} {\bibinfo
  {title} {Experimental identification of quantum spin liquids},\ }\href@noop
  {} {\bibfield  {journal} {\bibinfo  {journal} {npj Quantum Mater.}\ }\textbf
  {\bibinfo {volume} {4}},\ \bibinfo {pages} {12} (\bibinfo {year}
  {2019})}\BibitemShut {NoStop}%
\bibitem [{\citenamefont {Chen}(2019)}]{PhysRevResearch.1.033141}%
  \BibitemOpen
  \bibfield  {author} {\bibinfo {author} {\bibfnamefont {G.}~\bibnamefont
  {Chen}},\ }\bibfield  {title} {\bibinfo {title} {Intrinsic transverse field
  in frustrated quantum ising magnets: Physical origin and quantum effects},\
  }\href {https://doi.org/10.1103/PhysRevResearch.1.033141} {\bibfield
  {journal} {\bibinfo  {journal} {Phys. Rev. Res.}\ }\textbf {\bibinfo {volume}
  {1}},\ \bibinfo {pages} {033141} (\bibinfo {year} {2019})}\BibitemShut
  {NoStop}%
\bibitem [{\citenamefont {Liu}\ \emph {et~al.}(2020{\natexlab{a}})\citenamefont
  {Liu}, \citenamefont {Huang},\ and\ \citenamefont
  {Chen}}]{PhysRevResearch.2.043013}%
  \BibitemOpen
  \bibfield  {author} {\bibinfo {author} {\bibfnamefont {C.}~\bibnamefont
  {Liu}}, \bibinfo {author} {\bibfnamefont {C.-J.}\ \bibnamefont {Huang}},\
  and\ \bibinfo {author} {\bibfnamefont {G.}~\bibnamefont {Chen}},\ }\bibfield
  {title} {\bibinfo {title} {Intrinsic quantum ising model on a triangular
  lattice magnet {$\rm TmMgGaO_4$}},\ }\href
  {https://doi.org/10.1103/PhysRevResearch.2.043013} {\bibfield  {journal}
  {\bibinfo  {journal} {Phys. Rev. Res.}\ }\textbf {\bibinfo {volume} {2}},\
  \bibinfo {pages} {043013} (\bibinfo {year} {2020}{\natexlab{a}})}\BibitemShut
  {NoStop}%
\bibitem [{\citenamefont {Chubukov}\ and\ \citenamefont
  {Golosov}(1991)}]{AVChubukov_1991}%
  \BibitemOpen
  \bibfield  {author} {\bibinfo {author} {\bibfnamefont {A.~V.}\ \bibnamefont
  {Chubukov}}\ and\ \bibinfo {author} {\bibfnamefont {D.~I.}\ \bibnamefont
  {Golosov}},\ }\bibfield  {title} {\bibinfo {title} {Quantum theory of an
  antiferromagnet on a triangular lattice in a magnetic field},\ }\href
  {https://doi.org/10.1088/0953-8984/3/1/005} {\bibfield  {journal} {\bibinfo
  {journal} {J. Phys.:Condens. Matter}\ }\textbf {\bibinfo {volume} {3}},\
  \bibinfo {pages} {69} (\bibinfo {year} {1991})}\BibitemShut {NoStop}%
\bibitem [{\citenamefont {Bordelon}\ \emph {et~al.}(2019)\citenamefont
  {Bordelon}, \citenamefont {Kenney}, \citenamefont {Liu}, \citenamefont
  {Hogan}, \citenamefont {Posthuma}, \citenamefont {Kavand}, \citenamefont
  {Lyu}, \citenamefont {Sherwin}, \citenamefont {Butch}, \citenamefont {Brown}
  \emph {et~al.}}]{bordelon2019field}%
  \BibitemOpen
  \bibfield  {author} {\bibinfo {author} {\bibfnamefont {M.~M.}\ \bibnamefont
  {Bordelon}}, \bibinfo {author} {\bibfnamefont {E.}~\bibnamefont {Kenney}},
  \bibinfo {author} {\bibfnamefont {C.}~\bibnamefont {Liu}}, \bibinfo {author}
  {\bibfnamefont {T.}~\bibnamefont {Hogan}}, \bibinfo {author} {\bibfnamefont
  {L.}~\bibnamefont {Posthuma}}, \bibinfo {author} {\bibfnamefont
  {M.}~\bibnamefont {Kavand}}, \bibinfo {author} {\bibfnamefont
  {Y.}~\bibnamefont {Lyu}}, \bibinfo {author} {\bibfnamefont {M.}~\bibnamefont
  {Sherwin}}, \bibinfo {author} {\bibfnamefont {N.~P.}\ \bibnamefont {Butch}},
  \bibinfo {author} {\bibfnamefont {C.}~\bibnamefont {Brown}}, \emph {et~al.},\
  }\bibfield  {title} {\bibinfo {title} {Field-tunable quantum disordered
  ground state in the triangular-lattice antiferromagnet {$\rm NaYbO_2$}},\
  }\href@noop {} {\bibfield  {journal} {\bibinfo  {journal} {Nat. Phys.}\
  }\textbf {\bibinfo {volume} {15}},\ \bibinfo {pages} {1058} (\bibinfo {year}
  {2019})}\BibitemShut {NoStop}%
\bibitem [{\citenamefont {Zhou}\ \emph {et~al.}(2012)\citenamefont {Zhou},
  \citenamefont {Xu}, \citenamefont {Hallas}, \citenamefont {Silverstein},
  \citenamefont {Wiebe}, \citenamefont {Umegaki}, \citenamefont {Yan},
  \citenamefont {Murphy}, \citenamefont {Park}, \citenamefont {Qiu},
  \citenamefont {Copley}, \citenamefont {Gardner},\ and\ \citenamefont
  {Takano}}]{PhysRevLett.109.267206}%
  \BibitemOpen
  \bibfield  {author} {\bibinfo {author} {\bibfnamefont {H.~D.}\ \bibnamefont
  {Zhou}}, \bibinfo {author} {\bibfnamefont {C.}~\bibnamefont {Xu}}, \bibinfo
  {author} {\bibfnamefont {A.~M.}\ \bibnamefont {Hallas}}, \bibinfo {author}
  {\bibfnamefont {H.~J.}\ \bibnamefont {Silverstein}}, \bibinfo {author}
  {\bibfnamefont {C.~R.}\ \bibnamefont {Wiebe}}, \bibinfo {author}
  {\bibfnamefont {I.}~\bibnamefont {Umegaki}}, \bibinfo {author} {\bibfnamefont
  {J.~Q.}\ \bibnamefont {Yan}}, \bibinfo {author} {\bibfnamefont {T.~P.}\
  \bibnamefont {Murphy}}, \bibinfo {author} {\bibfnamefont {J.-H.}\
  \bibnamefont {Park}}, \bibinfo {author} {\bibfnamefont {Y.}~\bibnamefont
  {Qiu}}, \bibinfo {author} {\bibfnamefont {J.~R.~D.}\ \bibnamefont {Copley}},
  \bibinfo {author} {\bibfnamefont {J.~S.}\ \bibnamefont {Gardner}},\ and\
  \bibinfo {author} {\bibfnamefont {Y.}~\bibnamefont {Takano}},\ }\bibfield
  {title} {\bibinfo {title} {Successive phase transitions and extended
  spin-excitation continuum in the $s\mathbf{=}\frac{1}{2}$ triangular-lattice
  antiferromagnet {$\rm Ba_3CoSb_2O_9$}},\ }\href
  {https://doi.org/10.1103/PhysRevLett.109.267206} {\bibfield  {journal}
  {\bibinfo  {journal} {Phys. Rev. Lett.}\ }\textbf {\bibinfo {volume} {109}},\
  \bibinfo {pages} {267206} (\bibinfo {year} {2012})}\BibitemShut {NoStop}%
\bibitem [{\citenamefont {Abragam}\ and\ \citenamefont
  {Pryce}(1951)}]{abragam1951theory}%
  \BibitemOpen
  \bibfield  {author} {\bibinfo {author} {\bibfnamefont {A.}~\bibnamefont
  {Abragam}}\ and\ \bibinfo {author} {\bibfnamefont {M.~H.~L.}\ \bibnamefont
  {Pryce}},\ }\bibfield  {title} {\bibinfo {title} {The theory of paramagnetic
  resonance in hydrated cobalt salts},\ }\href@noop {} {\bibfield  {journal}
  {\bibinfo  {journal} {Proc. R. Soc. London, Ser. A}\ }\textbf {\bibinfo
  {volume} {206}},\ \bibinfo {pages} {173} (\bibinfo {year}
  {1951})}\BibitemShut {NoStop}%
\bibitem [{\citenamefont {Lines}(1963)}]{PhysRev.131.546}%
  \BibitemOpen
  \bibfield  {author} {\bibinfo {author} {\bibfnamefont {M.~E.}\ \bibnamefont
  {Lines}},\ }\bibfield  {title} {\bibinfo {title} {Magnetic properties of
  {$\rm CoCl_2$} and {$\rm NiCl_2$}},\ }\href
  {https://doi.org/10.1103/PhysRev.131.546} {\bibfield  {journal} {\bibinfo
  {journal} {Phys. Rev.}\ }\textbf {\bibinfo {volume} {131}},\ \bibinfo {pages}
  {546} (\bibinfo {year} {1963})}\BibitemShut {NoStop}%
\bibitem [{\citenamefont {Shiba}\ \emph {et~al.}(2003)\citenamefont {Shiba},
  \citenamefont {Ueda}, \citenamefont {Okunishi}, \citenamefont {Kimura},\ and\
  \citenamefont {Kindo}}]{doi:10.1143/JPSJ.72.2326}%
  \BibitemOpen
  \bibfield  {author} {\bibinfo {author} {\bibfnamefont {H.}~\bibnamefont
  {Shiba}}, \bibinfo {author} {\bibfnamefont {Y.}~\bibnamefont {Ueda}},
  \bibinfo {author} {\bibfnamefont {K.}~\bibnamefont {Okunishi}}, \bibinfo
  {author} {\bibfnamefont {S.}~\bibnamefont {Kimura}},\ and\ \bibinfo {author}
  {\bibfnamefont {K.}~\bibnamefont {Kindo}},\ }\bibfield  {title} {\bibinfo
  {title} {Exchange interaction via crystal-field excited states and its
  importance in {$\rm CsCoCl_3$}},\ }\href
  {https://doi.org/10.1143/JPSJ.72.2326} {\bibfield  {journal} {\bibinfo
  {journal} {J. Phys. Soc. Jpn.}\ }\textbf {\bibinfo {volume} {72}},\ \bibinfo
  {pages} {2326} (\bibinfo {year} {2003})}\BibitemShut {NoStop}%
\bibitem [{\citenamefont {Chanlert}\ \emph {et~al.}(2016)\citenamefont
  {Chanlert}, \citenamefont {Kurita}, \citenamefont {Tanaka}, \citenamefont
  {Goto}, \citenamefont {Matsuo},\ and\ \citenamefont
  {Kindo}}]{PhysRevB.93.094420}%
  \BibitemOpen
  \bibfield  {author} {\bibinfo {author} {\bibfnamefont {P.}~\bibnamefont
  {Chanlert}}, \bibinfo {author} {\bibfnamefont {N.}~\bibnamefont {Kurita}},
  \bibinfo {author} {\bibfnamefont {H.}~\bibnamefont {Tanaka}}, \bibinfo
  {author} {\bibfnamefont {D.}~\bibnamefont {Goto}}, \bibinfo {author}
  {\bibfnamefont {A.}~\bibnamefont {Matsuo}},\ and\ \bibinfo {author}
  {\bibfnamefont {K.}~\bibnamefont {Kindo}},\ }\bibfield  {title} {\bibinfo
  {title} {Field-driven successive phase transitions in the
  quasi-two-dimensional frustrated antiferromagnet {$\rm Ba_2CoTeO_6$} and
  highly degenerate classical ground states},\ }\href
  {https://doi.org/10.1103/PhysRevB.93.094420} {\bibfield  {journal} {\bibinfo
  {journal} {Phys. Rev. B}\ }\textbf {\bibinfo {volume} {93}},\ \bibinfo
  {pages} {094420} (\bibinfo {year} {2016})}\BibitemShut {NoStop}%
\bibitem [{\citenamefont {Ivanov}\ \emph {et~al.}(2010)\citenamefont {Ivanov},
  \citenamefont {Nordblad}, \citenamefont {Mathieu}, \citenamefont {Tellgren},\
  and\ \citenamefont {Ritter}}]{B927498G}%
  \BibitemOpen
  \bibfield  {author} {\bibinfo {author} {\bibfnamefont {S.~A.}\ \bibnamefont
  {Ivanov}}, \bibinfo {author} {\bibfnamefont {P.}~\bibnamefont {Nordblad}},
  \bibinfo {author} {\bibfnamefont {R.}~\bibnamefont {Mathieu}}, \bibinfo
  {author} {\bibfnamefont {R.}~\bibnamefont {Tellgren}},\ and\ \bibinfo
  {author} {\bibfnamefont {C.}~\bibnamefont {Ritter}},\ }\bibfield  {title}
  {\bibinfo {title} {Neutron diffraction studies and the magnetism of an
  ordered perovskite: {$\rm Ba_2CoTeO_6$}},\ }\href
  {https://doi.org/10.1039/B927498G} {\bibfield  {journal} {\bibinfo  {journal}
  {Dalton Trans.}\ }\textbf {\bibinfo {volume} {39}},\ \bibinfo {pages} {5490}
  (\bibinfo {year} {2010})}\BibitemShut {NoStop}%
\bibitem [{\citenamefont {Susuki}\ \emph {et~al.}(2013)\citenamefont {Susuki},
  \citenamefont {Kurita}, \citenamefont {Tanaka}, \citenamefont {Nojiri},
  \citenamefont {Matsuo}, \citenamefont {Kindo},\ and\ \citenamefont
  {Tanaka}}]{PhysRevLett.110.267201}%
  \BibitemOpen
  \bibfield  {author} {\bibinfo {author} {\bibfnamefont {T.}~\bibnamefont
  {Susuki}}, \bibinfo {author} {\bibfnamefont {N.}~\bibnamefont {Kurita}},
  \bibinfo {author} {\bibfnamefont {T.}~\bibnamefont {Tanaka}}, \bibinfo
  {author} {\bibfnamefont {H.}~\bibnamefont {Nojiri}}, \bibinfo {author}
  {\bibfnamefont {A.}~\bibnamefont {Matsuo}}, \bibinfo {author} {\bibfnamefont
  {K.}~\bibnamefont {Kindo}},\ and\ \bibinfo {author} {\bibfnamefont
  {H.}~\bibnamefont {Tanaka}},\ }\bibfield  {title} {\bibinfo {title}
  {Magnetization process and collective excitations in the $s\mathbf{=}1/2$
  triangular-lattice heisenberg antiferromagnet {$\rm Ba_3CoSb_2O_9$}},\ }\href
  {https://doi.org/10.1103/PhysRevLett.110.267201} {\bibfield  {journal}
  {\bibinfo  {journal} {Phys. Rev. Lett.}\ }\textbf {\bibinfo {volume} {110}},\
  \bibinfo {pages} {267201} (\bibinfo {year} {2013})}\BibitemShut {NoStop}%
\bibitem [{\citenamefont {Ma}\ \emph {et~al.}(2016)\citenamefont {Ma},
  \citenamefont {Kamiya}, \citenamefont {Hong}, \citenamefont {Cao},
  \citenamefont {Ehlers}, \citenamefont {Tian}, \citenamefont {Batista},
  \citenamefont {Dun}, \citenamefont {Zhou},\ and\ \citenamefont
  {Matsuda}}]{PhysRevLett.116.087201}%
  \BibitemOpen
  \bibfield  {author} {\bibinfo {author} {\bibfnamefont {J.}~\bibnamefont
  {Ma}}, \bibinfo {author} {\bibfnamefont {Y.}~\bibnamefont {Kamiya}}, \bibinfo
  {author} {\bibfnamefont {T.}~\bibnamefont {Hong}}, \bibinfo {author}
  {\bibfnamefont {H.~B.}\ \bibnamefont {Cao}}, \bibinfo {author} {\bibfnamefont
  {G.}~\bibnamefont {Ehlers}}, \bibinfo {author} {\bibfnamefont
  {W.}~\bibnamefont {Tian}}, \bibinfo {author} {\bibfnamefont {C.~D.}\
  \bibnamefont {Batista}}, \bibinfo {author} {\bibfnamefont {Z.~L.}\
  \bibnamefont {Dun}}, \bibinfo {author} {\bibfnamefont {H.~D.}\ \bibnamefont
  {Zhou}},\ and\ \bibinfo {author} {\bibfnamefont {M.}~\bibnamefont
  {Matsuda}},\ }\bibfield  {title} {\bibinfo {title} {Static and dynamical
  properties of the spin-$1/2$ equilateral triangular-lattice antiferromagnet
  {$\rm Ba_3CoSb_2O_9$}},\ }\href
  {https://doi.org/10.1103/PhysRevLett.116.087201} {\bibfield  {journal}
  {\bibinfo  {journal} {Phys. Rev. Lett.}\ }\textbf {\bibinfo {volume} {116}},\
  \bibinfo {pages} {087201} (\bibinfo {year} {2016})}\BibitemShut {NoStop}%
\bibitem [{\citenamefont {Ito}\ \emph {et~al.}(2017)\citenamefont {Ito},
  \citenamefont {Kurita}, \citenamefont {Tanaka}, \citenamefont
  {Ohira-Kawamura}, \citenamefont {Nakajima}, \citenamefont {Itoh},
  \citenamefont {Kuwahara},\ and\ \citenamefont {Kakurai}}]{ito2017structure}%
  \BibitemOpen
  \bibfield  {author} {\bibinfo {author} {\bibfnamefont {S.}~\bibnamefont
  {Ito}}, \bibinfo {author} {\bibfnamefont {N.}~\bibnamefont {Kurita}},
  \bibinfo {author} {\bibfnamefont {H.}~\bibnamefont {Tanaka}}, \bibinfo
  {author} {\bibfnamefont {S.}~\bibnamefont {Ohira-Kawamura}}, \bibinfo
  {author} {\bibfnamefont {K.}~\bibnamefont {Nakajima}}, \bibinfo {author}
  {\bibfnamefont {S.}~\bibnamefont {Itoh}}, \bibinfo {author} {\bibfnamefont
  {K.}~\bibnamefont {Kuwahara}},\ and\ \bibinfo {author} {\bibfnamefont
  {K.}~\bibnamefont {Kakurai}},\ }\bibfield  {title} {\bibinfo {title}
  {Structure of the magnetic excitations in the spin-1/2 triangular-lattice
  heisenberg antiferromagnet {$\rm Ba_3CoSb_2O_9$}},\ }\href@noop {} {\bibfield
   {journal} {\bibinfo  {journal} {Nat. Commun.}\ }\textbf {\bibinfo {volume}
  {8}},\ \bibinfo {pages} {235} (\bibinfo {year} {2017})}\BibitemShut {NoStop}%
\bibitem [{\citenamefont {Kamiya}\ \emph {et~al.}(2018)\citenamefont {Kamiya},
  \citenamefont {Ge}, \citenamefont {Hong}, \citenamefont {Qiu}, \citenamefont
  {Quintero-Castro}, \citenamefont {Lu}, \citenamefont {Cao}, \citenamefont
  {Matsuda}, \citenamefont {Choi}, \citenamefont {Batista} \emph
  {et~al.}}]{kamiya2018nature}%
  \BibitemOpen
  \bibfield  {author} {\bibinfo {author} {\bibfnamefont {Y.}~\bibnamefont
  {Kamiya}}, \bibinfo {author} {\bibfnamefont {L.}~\bibnamefont {Ge}}, \bibinfo
  {author} {\bibfnamefont {T.}~\bibnamefont {Hong}}, \bibinfo {author}
  {\bibfnamefont {Y.}~\bibnamefont {Qiu}}, \bibinfo {author} {\bibfnamefont
  {D.}~\bibnamefont {Quintero-Castro}}, \bibinfo {author} {\bibfnamefont
  {Z.}~\bibnamefont {Lu}}, \bibinfo {author} {\bibfnamefont {H.}~\bibnamefont
  {Cao}}, \bibinfo {author} {\bibfnamefont {M.}~\bibnamefont {Matsuda}},
  \bibinfo {author} {\bibfnamefont {E.}~\bibnamefont {Choi}}, \bibinfo {author}
  {\bibfnamefont {C.}~\bibnamefont {Batista}}, \emph {et~al.},\ }\bibfield
  {title} {\bibinfo {title} {The nature of spin excitations in the one-third
  magnetization plateau phase of {$\rm Ba_3CoSb_2O_9$}},\ }\href@noop {}
  {\bibfield  {journal} {\bibinfo  {journal} {Nat. Commun.}\ }\textbf {\bibinfo
  {volume} {9}},\ \bibinfo {pages} {2666} (\bibinfo {year} {2018})}\BibitemShut
  {NoStop}%
\bibitem [{\citenamefont {Kojima}\ \emph {et~al.}(2018)\citenamefont {Kojima},
  \citenamefont {Watanabe}, \citenamefont {Kurita}, \citenamefont {Tanaka},
  \citenamefont {Matsuo}, \citenamefont {Kindo},\ and\ \citenamefont
  {Avdeev}}]{kojima2018quantum}%
  \BibitemOpen
  \bibfield  {author} {\bibinfo {author} {\bibfnamefont {Y.}~\bibnamefont
  {Kojima}}, \bibinfo {author} {\bibfnamefont {M.}~\bibnamefont {Watanabe}},
  \bibinfo {author} {\bibfnamefont {N.}~\bibnamefont {Kurita}}, \bibinfo
  {author} {\bibfnamefont {H.}~\bibnamefont {Tanaka}}, \bibinfo {author}
  {\bibfnamefont {A.}~\bibnamefont {Matsuo}}, \bibinfo {author} {\bibfnamefont
  {K.}~\bibnamefont {Kindo}},\ and\ \bibinfo {author} {\bibfnamefont
  {M.}~\bibnamefont {Avdeev}},\ }\bibfield  {title} {\bibinfo {title} {Quantum
  magnetic properties of the spin-1/2 triangular-lattice antiferromagnet {$\rm
  Ba_2La_2CoTe_2O_{12}$}},\ }\href@noop {} {\bibfield  {journal} {\bibinfo
  {journal} {Phys. Rev. B}\ }\textbf {\bibinfo {volume} {98}},\ \bibinfo
  {pages} {174406} (\bibinfo {year} {2018})}\BibitemShut {NoStop}%
\bibitem [{\citenamefont {Park}\ \emph {et~al.}(2024)\citenamefont {Park},
  \citenamefont {Ghioldi}, \citenamefont {May}, \citenamefont {Kolopus},
  \citenamefont {Podlesnyak}, \citenamefont {Calder}, \citenamefont {Paddison},
  \citenamefont {Trumper}, \citenamefont {Manuel}, \citenamefont {Batista}
  \emph {et~al.}}]{park2024anomalous}%
  \BibitemOpen
  \bibfield  {author} {\bibinfo {author} {\bibfnamefont {P.}~\bibnamefont
  {Park}}, \bibinfo {author} {\bibfnamefont {E.~A.}\ \bibnamefont {Ghioldi}},
  \bibinfo {author} {\bibfnamefont {A.~F.}\ \bibnamefont {May}}, \bibinfo
  {author} {\bibfnamefont {J.~A.}\ \bibnamefont {Kolopus}}, \bibinfo {author}
  {\bibfnamefont {A.~A.}\ \bibnamefont {Podlesnyak}}, \bibinfo {author}
  {\bibfnamefont {S.}~\bibnamefont {Calder}}, \bibinfo {author} {\bibfnamefont
  {J.~A.}\ \bibnamefont {Paddison}}, \bibinfo {author} {\bibfnamefont {A.~E.}\
  \bibnamefont {Trumper}}, \bibinfo {author} {\bibfnamefont {L.~O.}\
  \bibnamefont {Manuel}}, \bibinfo {author} {\bibfnamefont {C.~D.}\
  \bibnamefont {Batista}}, \emph {et~al.},\ }\bibfield  {title} {\bibinfo
  {title} {Anomalous continuum scattering and higher-order van hove singularity
  in the strongly anisotropic s= 1/2 triangular lattice antiferromagnet},\
  }\href@noop {} {\bibfield  {journal} {\bibinfo  {journal} {Nat. Commun.}\
  }\textbf {\bibinfo {volume} {15}},\ \bibinfo {pages} {7264} (\bibinfo {year}
  {2024})}\BibitemShut {NoStop}%
\bibitem [{\citenamefont {Zhong}\ \emph
  {et~al.}(2020{\natexlab{a}})\citenamefont {Zhong}, \citenamefont {Guo},\ and\
  \citenamefont {Cava}}]{PhysRevMaterials.4.084406}%
  \BibitemOpen
  \bibfield  {author} {\bibinfo {author} {\bibfnamefont {R.}~\bibnamefont
  {Zhong}}, \bibinfo {author} {\bibfnamefont {S.}~\bibnamefont {Guo}},\ and\
  \bibinfo {author} {\bibfnamefont {R.~J.}\ \bibnamefont {Cava}},\ }\bibfield
  {title} {\bibinfo {title} {Frustrated magnetism in the layered triangular
  lattice materials {$\rm K_2Co(SeO_3)_2$} and {$\rm Rb_2Co(SeO_3)_2$}},\
  }\href {https://doi.org/10.1103/PhysRevMaterials.4.084406} {\bibfield
  {journal} {\bibinfo  {journal} {Phys. Rev. Mater.}\ }\textbf {\bibinfo
  {volume} {4}},\ \bibinfo {pages} {084406} (\bibinfo {year}
  {2020}{\natexlab{a}})}\BibitemShut {NoStop}%
\bibitem [{\citenamefont {Zhu}\ \emph {et~al.}(2025)\citenamefont {Zhu},
  \citenamefont {Chinellato}, \citenamefont {Romerio}, \citenamefont {Murai},
  \citenamefont {Ohira-Kawamura}, \citenamefont {Balz}, \citenamefont {Yan},
  \citenamefont {Gvasaliya}, \citenamefont {Kato}, \citenamefont {Batista}
  \emph {et~al.}}]{zhu2025wannier}%
  \BibitemOpen
  \bibfield  {author} {\bibinfo {author} {\bibfnamefont {M.}~\bibnamefont
  {Zhu}}, \bibinfo {author} {\bibfnamefont {L.~M.}\ \bibnamefont {Chinellato}},
  \bibinfo {author} {\bibfnamefont {V.}~\bibnamefont {Romerio}}, \bibinfo
  {author} {\bibfnamefont {N.}~\bibnamefont {Murai}}, \bibinfo {author}
  {\bibfnamefont {S.}~\bibnamefont {Ohira-Kawamura}}, \bibinfo {author}
  {\bibfnamefont {C.}~\bibnamefont {Balz}}, \bibinfo {author} {\bibfnamefont
  {Z.}~\bibnamefont {Yan}}, \bibinfo {author} {\bibfnamefont {S.}~\bibnamefont
  {Gvasaliya}}, \bibinfo {author} {\bibfnamefont {Y.}~\bibnamefont {Kato}},
  \bibinfo {author} {\bibfnamefont {C.}~\bibnamefont {Batista}}, \emph
  {et~al.},\ }\bibfield  {title} {\bibinfo {title} {Wannier states and spin
  supersolid physics in the triangular antiferromagnet {$\rm
  K_2Co(SeO_3)_2$}},\ }\href@noop {} {\bibfield  {journal} {\bibinfo  {journal}
  {npj Quantum Mater.}\ }\textbf {\bibinfo {volume} {10}},\ \bibinfo {pages}
  {74} (\bibinfo {year} {2025})}\BibitemShut {NoStop}%
\bibitem [{\citenamefont {Zhu}\ \emph {et~al.}(2024)\citenamefont {Zhu},
  \citenamefont {Romerio}, \citenamefont {Steiger}, \citenamefont {Nabi},
  \citenamefont {Murai}, \citenamefont {Ohira-Kawamura}, \citenamefont
  {Povarov}, \citenamefont {Skourski}, \citenamefont {Sibille}, \citenamefont
  {Keller} \emph {et~al.}}]{zhu2024continuum}%
  \BibitemOpen
  \bibfield  {author} {\bibinfo {author} {\bibfnamefont {M.}~\bibnamefont
  {Zhu}}, \bibinfo {author} {\bibfnamefont {V.}~\bibnamefont {Romerio}},
  \bibinfo {author} {\bibfnamefont {N.}~\bibnamefont {Steiger}}, \bibinfo
  {author} {\bibfnamefont {S.}~\bibnamefont {Nabi}}, \bibinfo {author}
  {\bibfnamefont {N.}~\bibnamefont {Murai}}, \bibinfo {author} {\bibfnamefont
  {S.}~\bibnamefont {Ohira-Kawamura}}, \bibinfo {author} {\bibfnamefont
  {K.~Y.}\ \bibnamefont {Povarov}}, \bibinfo {author} {\bibfnamefont
  {Y.}~\bibnamefont {Skourski}}, \bibinfo {author} {\bibfnamefont
  {R.}~\bibnamefont {Sibille}}, \bibinfo {author} {\bibfnamefont
  {L.}~\bibnamefont {Keller}}, \emph {et~al.},\ }\bibfield  {title} {\bibinfo
  {title} {Continuum excitations in a spin supersolid on a triangular
  lattice},\ }\href@noop {} {\bibfield  {journal} {\bibinfo  {journal} {Phys.
  Rev. Lett.}\ }\textbf {\bibinfo {volume} {133}},\ \bibinfo {pages} {186704}
  (\bibinfo {year} {2024})}\BibitemShut {NoStop}%
\bibitem [{\citenamefont {Mauri}\ and\ \citenamefont
  {Mila}(2025)}]{PhysRevB.111.L180402}%
  \BibitemOpen
  \bibfield  {author} {\bibinfo {author} {\bibfnamefont {A.}~\bibnamefont
  {Mauri}}\ and\ \bibinfo {author} {\bibfnamefont {F.}~\bibnamefont {Mila}},\
  }\bibfield  {title} {\bibinfo {title} {Slow convergence of spin-wave
  expansion and magnon dispersion in the 1/3 plateau of the triangular {XXZ}
  antiferromagnet},\ }\href {https://doi.org/10.1103/PhysRevB.111.L180402}
  {\bibfield  {journal} {\bibinfo  {journal} {Phys. Rev. B}\ }\textbf {\bibinfo
  {volume} {111}},\ \bibinfo {pages} {L180402} (\bibinfo {year}
  {2025})}\BibitemShut {NoStop}%
\bibitem [{\citenamefont {Shi}\ \emph {et~al.}(2025)\citenamefont {Shi},
  \citenamefont {Han}, \citenamefont {Yu}, \citenamefont {Ling}, \citenamefont
  {Tong}, \citenamefont {Xi}, \citenamefont {Shang}, \citenamefont {Wang},
  \citenamefont {Pi},\ and\ \citenamefont
  {Ma}}]{shi2025absencehighfieldspinsupersolid}%
  \BibitemOpen
  \bibfield  {author} {\bibinfo {author} {\bibfnamefont {K.}~\bibnamefont
  {Shi}}, \bibinfo {author} {\bibfnamefont {Y.~Q.}\ \bibnamefont {Han}},
  \bibinfo {author} {\bibfnamefont {B.~C.}\ \bibnamefont {Yu}}, \bibinfo
  {author} {\bibfnamefont {L.~S.}\ \bibnamefont {Ling}}, \bibinfo {author}
  {\bibfnamefont {W.}~\bibnamefont {Tong}}, \bibinfo {author} {\bibfnamefont
  {C.~Y.}\ \bibnamefont {Xi}}, \bibinfo {author} {\bibfnamefont
  {T.}~\bibnamefont {Shang}}, \bibinfo {author} {\bibfnamefont
  {Z.}~\bibnamefont {Wang}}, \bibinfo {author} {\bibfnamefont {L.}~\bibnamefont
  {Pi}},\ and\ \bibinfo {author} {\bibfnamefont {L.}~\bibnamefont {Ma}},\
  }\href {https://arxiv.org/abs/2509.06281} {\bibinfo {title} {Absence of
  high-field spin supersolid phase in {$\rm Rb_2Co(SeO_3)_2$} with a triangular
  lattice}} (\bibinfo {year} {2025}),\ \Eprint
  {https://arxiv.org/abs/2509.06281} {arXiv:2509.06281 [cond-mat.str-el]}
  \BibitemShut {NoStop}%
\bibitem [{\citenamefont {Xu}\ \emph {et~al.}(2024)\citenamefont {Xu},
  \citenamefont {Chen}, \citenamefont {Wang}, \citenamefont {Xie},
  \citenamefont {Broholm},\ and\ \citenamefont {Cava}}]{xu2024frustrated}%
  \BibitemOpen
  \bibfield  {author} {\bibinfo {author} {\bibfnamefont {X.}~\bibnamefont
  {Xu}}, \bibinfo {author} {\bibfnamefont {T.}~\bibnamefont {Chen}}, \bibinfo
  {author} {\bibfnamefont {H.}~\bibnamefont {Wang}}, \bibinfo {author}
  {\bibfnamefont {W.}~\bibnamefont {Xie}}, \bibinfo {author} {\bibfnamefont
  {C.}~\bibnamefont {Broholm}},\ and\ \bibinfo {author} {\bibfnamefont {R.~J.}\
  \bibnamefont {Cava}},\ }\bibfield  {title} {\bibinfo {title} {Frustrated
  magnetism in a potential quantum material based on spin-1/2 {$\rm Co^{2+}$}
  dimers},\ }\href@noop {} {\bibfield  {journal} {\bibinfo  {journal} {Chem.
  Mater.}\ }\textbf {\bibinfo {volume} {36}},\ \bibinfo {pages} {4157}
  (\bibinfo {year} {2024})}\BibitemShut {NoStop}%
\bibitem [{\citenamefont {Chen}(2024)}]{PhysRevLett.133.136703}%
  \BibitemOpen
  \bibfield  {author} {\bibinfo {author} {\bibfnamefont {G.~V.}\ \bibnamefont
  {Chen}},\ }\bibfield  {title} {\bibinfo {title} {Emergent
  {Berezinskii-Kosterlitz-Thouless} and {Kugel-Khomskii} physics in the
  triangular lattice bilayer colbaltate},\ }\href
  {https://doi.org/10.1103/PhysRevLett.133.136703} {\bibfield  {journal}
  {\bibinfo  {journal} {Phys. Rev. Lett.}\ }\textbf {\bibinfo {volume} {133}},\
  \bibinfo {pages} {136703} (\bibinfo {year} {2024})}\BibitemShut {NoStop}%
\bibitem [{\citenamefont {Zhong}\ \emph
  {et~al.}(2020{\natexlab{b}})\citenamefont {Zhong}, \citenamefont {Guo},
  \citenamefont {Nguyen},\ and\ \citenamefont {Cava}}]{PhysRevB.102.224430}%
  \BibitemOpen
  \bibfield  {author} {\bibinfo {author} {\bibfnamefont {R.}~\bibnamefont
  {Zhong}}, \bibinfo {author} {\bibfnamefont {S.}~\bibnamefont {Guo}}, \bibinfo
  {author} {\bibfnamefont {L.~T.}\ \bibnamefont {Nguyen}},\ and\ \bibinfo
  {author} {\bibfnamefont {R.~J.}\ \bibnamefont {Cava}},\ }\bibfield  {title}
  {\bibinfo {title} {Frustrated spin-1/2 dimer compound {$\rm
  K_2Co_2(SeO_3)_3$} with easy-axis anisotropy},\ }\href
  {https://doi.org/10.1103/PhysRevB.102.224430} {\bibfield  {journal} {\bibinfo
   {journal} {Phys. Rev. B}\ }\textbf {\bibinfo {volume} {102}},\ \bibinfo
  {pages} {224430} (\bibinfo {year} {2020}{\natexlab{b}})}\BibitemShut
  {NoStop}%
\bibitem [{\citenamefont {Zhong}\ \emph {et~al.}(2019)\citenamefont {Zhong},
  \citenamefont {Guo}, \citenamefont {Xu}, \citenamefont {Xu},\ and\
  \citenamefont {Cava}}]{doi:10.1073/pnas.1906483116}%
  \BibitemOpen
  \bibfield  {author} {\bibinfo {author} {\bibfnamefont {R.}~\bibnamefont
  {Zhong}}, \bibinfo {author} {\bibfnamefont {S.}~\bibnamefont {Guo}}, \bibinfo
  {author} {\bibfnamefont {G.}~\bibnamefont {Xu}}, \bibinfo {author}
  {\bibfnamefont {Z.}~\bibnamefont {Xu}},\ and\ \bibinfo {author}
  {\bibfnamefont {R.~J.}\ \bibnamefont {Cava}},\ }\bibfield  {title} {\bibinfo
  {title} {Strong quantum fluctuations in a quantum spin liquid candidate with
  a {Co-based} triangular lattice},\ }\href
  {https://doi.org/10.1073/pnas.1906483116} {\bibfield  {journal} {\bibinfo
  {journal} {Proc. Natl. Acad. Sci.}\ }\textbf {\bibinfo {volume} {116}},\
  \bibinfo {pages} {14505} (\bibinfo {year} {2019})}\BibitemShut {NoStop}%
\bibitem [{\citenamefont {Xiang}\ \emph {et~al.}(2024)\citenamefont {Xiang},
  \citenamefont {Zhang}, \citenamefont {Gao}, \citenamefont {Schmidt},
  \citenamefont {Schmalzl}, \citenamefont {Wang}, \citenamefont {Li},
  \citenamefont {Xi}, \citenamefont {Liu}, \citenamefont {Jin} \emph
  {et~al.}}]{xiang2024giant}%
  \BibitemOpen
  \bibfield  {author} {\bibinfo {author} {\bibfnamefont {J.}~\bibnamefont
  {Xiang}}, \bibinfo {author} {\bibfnamefont {C.}~\bibnamefont {Zhang}},
  \bibinfo {author} {\bibfnamefont {Y.}~\bibnamefont {Gao}}, \bibinfo {author}
  {\bibfnamefont {W.}~\bibnamefont {Schmidt}}, \bibinfo {author} {\bibfnamefont
  {K.}~\bibnamefont {Schmalzl}}, \bibinfo {author} {\bibfnamefont {C.-W.}\
  \bibnamefont {Wang}}, \bibinfo {author} {\bibfnamefont {B.}~\bibnamefont
  {Li}}, \bibinfo {author} {\bibfnamefont {N.}~\bibnamefont {Xi}}, \bibinfo
  {author} {\bibfnamefont {X.-Y.}\ \bibnamefont {Liu}}, \bibinfo {author}
  {\bibfnamefont {H.}~\bibnamefont {Jin}}, \emph {et~al.},\ }\bibfield  {title}
  {\bibinfo {title} {Giant magnetocaloric effect in spin supersolid candidate
  {$\rm Na_2BaCo(PO_4)_2$}},\ }\href@noop {} {\bibfield  {journal} {\bibinfo
  {journal} {Nature}\ }\textbf {\bibinfo {volume} {625}},\ \bibinfo {pages}
  {270} (\bibinfo {year} {2024})}\BibitemShut {NoStop}%
\bibitem [{\citenamefont {Gao}\ \emph {et~al.}(2022)\citenamefont {Gao},
  \citenamefont {Fan}, \citenamefont {Li}, \citenamefont {Yang}, \citenamefont
  {Zeng}, \citenamefont {Sheng}, \citenamefont {Zhong}, \citenamefont {Qi},
  \citenamefont {Wan},\ and\ \citenamefont {Li}}]{gao2022spin}%
  \BibitemOpen
  \bibfield  {author} {\bibinfo {author} {\bibfnamefont {Y.}~\bibnamefont
  {Gao}}, \bibinfo {author} {\bibfnamefont {Y.-C.}\ \bibnamefont {Fan}},
  \bibinfo {author} {\bibfnamefont {H.}~\bibnamefont {Li}}, \bibinfo {author}
  {\bibfnamefont {F.}~\bibnamefont {Yang}}, \bibinfo {author} {\bibfnamefont
  {X.-T.}\ \bibnamefont {Zeng}}, \bibinfo {author} {\bibfnamefont {X.-L.}\
  \bibnamefont {Sheng}}, \bibinfo {author} {\bibfnamefont {R.}~\bibnamefont
  {Zhong}}, \bibinfo {author} {\bibfnamefont {Y.}~\bibnamefont {Qi}}, \bibinfo
  {author} {\bibfnamefont {Y.}~\bibnamefont {Wan}},\ and\ \bibinfo {author}
  {\bibfnamefont {W.}~\bibnamefont {Li}},\ }\bibfield  {title} {\bibinfo
  {title} {Spin supersolidity in nearly ideal easy-axis triangular quantum
  antiferromagnet {$\rm Na_2BaCo(PO_4)_2$}},\ }\href@noop {} {\bibfield
  {journal} {\bibinfo  {journal} {npj Quantum Mater.}\ }\textbf {\bibinfo
  {volume} {7}},\ \bibinfo {pages} {89} (\bibinfo {year} {2022})}\BibitemShut
  {NoStop}%
\bibitem [{\citenamefont {Zhang}\ \emph {et~al.}(2022)\citenamefont {Zhang},
  \citenamefont {Xu}, \citenamefont {Zeng}, \citenamefont {Lyu}, \citenamefont
  {Lin}, \citenamefont {Hao}, \citenamefont {Deng}, \citenamefont {He},
  \citenamefont {Xiao}, \citenamefont {Ye} \emph {et~al.}}]{zhang2022doping}%
  \BibitemOpen
  \bibfield  {author} {\bibinfo {author} {\bibfnamefont {C.}~\bibnamefont
  {Zhang}}, \bibinfo {author} {\bibfnamefont {Q.}~\bibnamefont {Xu}}, \bibinfo
  {author} {\bibfnamefont {X.-T.}\ \bibnamefont {Zeng}}, \bibinfo {author}
  {\bibfnamefont {C.}~\bibnamefont {Lyu}}, \bibinfo {author} {\bibfnamefont
  {Z.}~\bibnamefont {Lin}}, \bibinfo {author} {\bibfnamefont {J.}~\bibnamefont
  {Hao}}, \bibinfo {author} {\bibfnamefont {S.}~\bibnamefont {Deng}}, \bibinfo
  {author} {\bibfnamefont {L.}~\bibnamefont {He}}, \bibinfo {author}
  {\bibfnamefont {Y.}~\bibnamefont {Xiao}}, \bibinfo {author} {\bibfnamefont
  {Y.}~\bibnamefont {Ye}}, \emph {et~al.},\ }\bibfield  {title} {\bibinfo
  {title} {Doping-induced structural transformation in the spin-1/2
  triangular-lattice antiferromagnet {$\rm Na_2Ba_{1-x}Sr_xCo(PO_4)_2$}},\
  }\href@noop {} {\bibfield  {journal} {\bibinfo  {journal} {J. Alloys Compd.}\
  }\textbf {\bibinfo {volume} {905}},\ \bibinfo {pages} {164147} (\bibinfo
  {year} {2022})}\BibitemShut {NoStop}%
\bibitem [{\citenamefont {Bader}\ \emph {et~al.}(2022)\citenamefont {Bader},
  \citenamefont {Langmann}, \citenamefont {Gegenwart},\ and\ \citenamefont
  {Tsirlin}}]{PhysRevB.106.054415}%
  \BibitemOpen
  \bibfield  {author} {\bibinfo {author} {\bibfnamefont {V.~P.}\ \bibnamefont
  {Bader}}, \bibinfo {author} {\bibfnamefont {J.}~\bibnamefont {Langmann}},
  \bibinfo {author} {\bibfnamefont {P.}~\bibnamefont {Gegenwart}},\ and\
  \bibinfo {author} {\bibfnamefont {A.~A.}\ \bibnamefont {Tsirlin}},\
  }\bibfield  {title} {\bibinfo {title} {Deformation of the triangular
  spin-$\frac{1}{2}$ lattice in
  {${\mathrm{Na}}_{2}{\mathrm{SrCo}(\mathrm{PO}}_{4}{)}_{2}$}},\ }\href
  {https://doi.org/10.1103/PhysRevB.106.054415} {\bibfield  {journal} {\bibinfo
   {journal} {Phys. Rev. B}\ }\textbf {\bibinfo {volume} {106}},\ \bibinfo
  {pages} {054415} (\bibinfo {year} {2022})}\BibitemShut {NoStop}%
\bibitem [{\citenamefont {Nakayama}\ \emph {et~al.}(2013)\citenamefont
  {Nakayama}, \citenamefont {Hara}, \citenamefont {Sato}, \citenamefont
  {Narumi},\ and\ \citenamefont {Nojiri}}]{Nakayama_2013}%
  \BibitemOpen
  \bibfield  {author} {\bibinfo {author} {\bibfnamefont {G.}~\bibnamefont
  {Nakayama}}, \bibinfo {author} {\bibfnamefont {S.}~\bibnamefont {Hara}},
  \bibinfo {author} {\bibfnamefont {H.}~\bibnamefont {Sato}}, \bibinfo {author}
  {\bibfnamefont {Y.}~\bibnamefont {Narumi}},\ and\ \bibinfo {author}
  {\bibfnamefont {H.}~\bibnamefont {Nojiri}},\ }\bibfield  {title} {\bibinfo
  {title} {Synthesis and magnetic properties of a new series of
  triangular-lattice magnets, {${\rm Na_2Ba}M{\rm V_2O_8}$ ($M$ = Ni, Co, and
  Mn)}},\ }\href {https://doi.org/10.1088/0953-8984/25/11/116003} {\bibfield
  {journal} {\bibinfo  {journal} {J. Phys.:Condens. Matter}\ }\textbf {\bibinfo
  {volume} {25}},\ \bibinfo {pages} {116003} (\bibinfo {year}
  {2013})}\BibitemShut {NoStop}%
\bibitem [{\citenamefont {M\"oller}\ \emph {et~al.}(2012)\citenamefont
  {M\"oller}, \citenamefont {Amuneke}, \citenamefont {Daniel}, \citenamefont
  {Lorenz}, \citenamefont {de~la Cruz}, \citenamefont {Gooch},\ and\
  \citenamefont {Chu}}]{PhysRevB.85.214422}%
  \BibitemOpen
  \bibfield  {author} {\bibinfo {author} {\bibfnamefont {A.}~\bibnamefont
  {M\"oller}}, \bibinfo {author} {\bibfnamefont {N.~E.}\ \bibnamefont
  {Amuneke}}, \bibinfo {author} {\bibfnamefont {P.}~\bibnamefont {Daniel}},
  \bibinfo {author} {\bibfnamefont {B.}~\bibnamefont {Lorenz}}, \bibinfo
  {author} {\bibfnamefont {C.~R.}\ \bibnamefont {de~la Cruz}}, \bibinfo
  {author} {\bibfnamefont {M.}~\bibnamefont {Gooch}},\ and\ \bibinfo {author}
  {\bibfnamefont {P.~C.~W.}\ \bibnamefont {Chu}},\ }\bibfield  {title}
  {\bibinfo {title} {{$A{\rm Ag_2}M{\rm [VO_4]_2}$} {($A={\rm Ba}, {\rm Sr}$;
  $M = {\rm Co}, {\rm Ni}$)}: A series of ferromagnetic insulators},\ }\href
  {https://doi.org/10.1103/PhysRevB.85.214422} {\bibfield  {journal} {\bibinfo
  {journal} {Phys. Rev. B}\ }\textbf {\bibinfo {volume} {85}},\ \bibinfo
  {pages} {214422} (\bibinfo {year} {2012})}\BibitemShut {NoStop}%
\bibitem [{\citenamefont {Sanjeewa}\ \emph {et~al.}(2017)\citenamefont
  {Sanjeewa}, \citenamefont {Garlea}, \citenamefont {McGuire}, \citenamefont
  {Frontzek}, \citenamefont {McMillen}, \citenamefont {Fulle},\ and\
  \citenamefont {Kolis}}]{doi:10.1021/acs.inorgchem.7b02024}%
  \BibitemOpen
  \bibfield  {author} {\bibinfo {author} {\bibfnamefont {L.~D.}\ \bibnamefont
  {Sanjeewa}}, \bibinfo {author} {\bibfnamefont {V.~O.}\ \bibnamefont
  {Garlea}}, \bibinfo {author} {\bibfnamefont {M.~A.}\ \bibnamefont {McGuire}},
  \bibinfo {author} {\bibfnamefont {M.}~\bibnamefont {Frontzek}}, \bibinfo
  {author} {\bibfnamefont {C.~D.}\ \bibnamefont {McMillen}}, \bibinfo {author}
  {\bibfnamefont {K.}~\bibnamefont {Fulle}},\ and\ \bibinfo {author}
  {\bibfnamefont {J.~W.}\ \bibnamefont {Kolis}},\ }\bibfield  {title} {\bibinfo
  {title} {Investigation of a structural phase transition and magnetic
  structure of {$\rm Na_2BaFe(VO_4)_2$}: A triangular magnetic lattice with a
  ferromagnetic ground state},\ }\href
  {https://doi.org/10.1021/acs.inorgchem.7b02024} {\bibfield  {journal}
  {\bibinfo  {journal} {Inorg. Chem.}\ }\textbf {\bibinfo {volume} {56}},\
  \bibinfo {pages} {14842} (\bibinfo {year} {2017})}\BibitemShut {NoStop}%
\bibitem [{\citenamefont {Sanjeewa}\ \emph {et~al.}(2016)\citenamefont
  {Sanjeewa}, \citenamefont {McMillen}, \citenamefont {Willett}, \citenamefont
  {Chumanov},\ and\ \citenamefont {Kolis}}]{SANJEEWA201661}%
  \BibitemOpen
  \bibfield  {author} {\bibinfo {author} {\bibfnamefont {L.~D.}\ \bibnamefont
  {Sanjeewa}}, \bibinfo {author} {\bibfnamefont {C.~D.}\ \bibnamefont
  {McMillen}}, \bibinfo {author} {\bibfnamefont {D.}~\bibnamefont {Willett}},
  \bibinfo {author} {\bibfnamefont {G.}~\bibnamefont {Chumanov}},\ and\
  \bibinfo {author} {\bibfnamefont {J.~W.}\ \bibnamefont {Kolis}},\ }\bibfield
  {title} {\bibinfo {title} {Hydrothermal synthesis of single crystals of
  transition metal vanadates in the glaserite phase},\ }\href
  {https://doi.org/https://doi.org/10.1016/j.jssc.2015.07.039} {\bibfield
  {journal} {\bibinfo  {journal} {J. Solid State Chem.}\ }\textbf {\bibinfo
  {volume} {236}},\ \bibinfo {pages} {61} (\bibinfo {year} {2016})}\BibitemShut
  {NoStop}%
\bibitem [{\citenamefont {Chen}\ \emph {et~al.}(2018)\citenamefont {Chen},
  \citenamefont {Kang}, \citenamefont {Lu}, \citenamefont {Luo}, \citenamefont
  {Wang},\ and\ \citenamefont {He}}]{CHEN2018370}%
  \BibitemOpen
  \bibfield  {author} {\bibinfo {author} {\bibfnamefont {J.}~\bibnamefont
  {Chen}}, \bibinfo {author} {\bibfnamefont {L.}~\bibnamefont {Kang}}, \bibinfo
  {author} {\bibfnamefont {H.}~\bibnamefont {Lu}}, \bibinfo {author}
  {\bibfnamefont {P.}~\bibnamefont {Luo}}, \bibinfo {author} {\bibfnamefont
  {F.}~\bibnamefont {Wang}},\ and\ \bibinfo {author} {\bibfnamefont
  {L.}~\bibnamefont {He}},\ }\bibfield  {title} {\bibinfo {title} {The general
  purpose powder diffractometer at csns},\ }\href
  {https://doi.org/https://doi.org/10.1016/j.physb.2017.11.005} {\bibfield
  {journal} {\bibinfo  {journal} {PHYSICA B}\ }\textbf {\bibinfo {volume}
  {551}},\ \bibinfo {pages} {370} (\bibinfo {year} {2018})}\BibitemShut
  {NoStop}%
\bibitem [{\citenamefont {Guo}\ \emph {et~al.}(2013)\citenamefont {Guo},
  \citenamefont {Han}, \citenamefont {Yuan}, \citenamefont {Sun}, \citenamefont
  {Liu}, \citenamefont {Chen}, \citenamefont {Yang},\ and\ \citenamefont
  {Du}}]{HGuo}%
  \BibitemOpen
  \bibfield  {author} {\bibinfo {author} {\bibfnamefont {H.}~\bibnamefont
  {Guo}}, \bibinfo {author} {\bibfnamefont {W.}~\bibnamefont {Han}}, \bibinfo
  {author} {\bibfnamefont {Z.}~\bibnamefont {Yuan}}, \bibinfo {author}
  {\bibfnamefont {K.}~\bibnamefont {Sun}}, \bibinfo {author} {\bibfnamefont
  {Y.}~\bibnamefont {Liu}}, \bibinfo {author} {\bibfnamefont {D.}~\bibnamefont
  {Chen}}, \bibinfo {author} {\bibfnamefont {J.}~\bibnamefont {Yang}},\ and\
  \bibinfo {author} {\bibfnamefont {H.}~\bibnamefont {Du}},\ }\bibfield
  {title} {\bibinfo {title} {Development of the high intensity powder neutron
  diffractometer at china advanced research reactor},\ }\href@noop {}
  {\bibfield  {journal} {\bibinfo  {journal} {Annu. Rep. China Inst. Atomic
  Energy}\ }\textbf {\bibinfo {volume} {1}},\ \bibinfo {pages} {92} (\bibinfo
  {year} {2013})}\BibitemShut {NoStop}%
\bibitem [{\citenamefont
  {Rodríguez-Carvajal}(1993)}]{RODRIGUEZCARVAJAL199355}%
  \BibitemOpen
  \bibfield  {author} {\bibinfo {author} {\bibfnamefont {J.}~\bibnamefont
  {Rodríguez-Carvajal}},\ }\bibfield  {title} {\bibinfo {title} {Recent
  advances in magnetic structure determination by neutron powder diffraction},\
  }\href {https://doi.org/https://doi.org/10.1016/0921-4526(93)90108-I}
  {\bibfield  {journal} {\bibinfo  {journal} {PHYSICA B}\ }\textbf {\bibinfo
  {volume} {192}},\ \bibinfo {pages} {55} (\bibinfo {year} {1993})}\BibitemShut
  {NoStop}%
\bibitem [{\citenamefont {Liu}\ \emph {et~al.}(2020{\natexlab{b}})\citenamefont
  {Liu}, \citenamefont {Chaloupka},\ and\ \citenamefont
  {Khaliullin}}]{PhysRevLett.125.047201}%
  \BibitemOpen
  \bibfield  {author} {\bibinfo {author} {\bibfnamefont {H.}~\bibnamefont
  {Liu}}, \bibinfo {author} {\bibfnamefont {J.~c.~v.}\ \bibnamefont
  {Chaloupka}},\ and\ \bibinfo {author} {\bibfnamefont {G.}~\bibnamefont
  {Khaliullin}},\ }\bibfield  {title} {\bibinfo {title} {Kitaev spin liquid in
  $3d$ transition metal compounds},\ }\href
  {https://doi.org/10.1103/PhysRevLett.125.047201} {\bibfield  {journal}
  {\bibinfo  {journal} {Phys. Rev. Lett.}\ }\textbf {\bibinfo {volume} {125}},\
  \bibinfo {pages} {047201} (\bibinfo {year} {2020}{\natexlab{b}})}\BibitemShut
  {NoStop}%
\bibitem [{\citenamefont {Popescu}\ \emph {et~al.}(2025)\citenamefont
  {Popescu}, \citenamefont {Gora}, \citenamefont {Demmel}, \citenamefont {Xu},
  \citenamefont {Zhong}, \citenamefont {Williams}, \citenamefont {Cava},
  \citenamefont {Xu},\ and\ \citenamefont {Stock}}]{PhysRevLett.134.136703}%
  \BibitemOpen
  \bibfield  {author} {\bibinfo {author} {\bibfnamefont {T.~I.}\ \bibnamefont
  {Popescu}}, \bibinfo {author} {\bibfnamefont {N.}~\bibnamefont {Gora}},
  \bibinfo {author} {\bibfnamefont {F.}~\bibnamefont {Demmel}}, \bibinfo
  {author} {\bibfnamefont {Z.}~\bibnamefont {Xu}}, \bibinfo {author}
  {\bibfnamefont {R.}~\bibnamefont {Zhong}}, \bibinfo {author} {\bibfnamefont
  {T.~J.}\ \bibnamefont {Williams}}, \bibinfo {author} {\bibfnamefont {R.~J.}\
  \bibnamefont {Cava}}, \bibinfo {author} {\bibfnamefont {G.}~\bibnamefont
  {Xu}},\ and\ \bibinfo {author} {\bibfnamefont {C.}~\bibnamefont {Stock}},\
  }\bibfield  {title} {\bibinfo {title} {Zeeman split kramers doublets in
  spin-supersolid candidate {$\rm Na_2BaCo(PO_4)_2$}},\ }\href
  {https://doi.org/10.1103/PhysRevLett.134.136703} {\bibfield  {journal}
  {\bibinfo  {journal} {Phys. Rev. Lett.}\ }\textbf {\bibinfo {volume} {134}},\
  \bibinfo {pages} {136703} (\bibinfo {year} {2025})}\BibitemShut {NoStop}%
\bibitem [{\citenamefont {Somesh}\ \emph {et~al.}(2021)\citenamefont {Somesh},
  \citenamefont {Furukawa}, \citenamefont {Simutis}, \citenamefont {Bert},
  \citenamefont {Prinz-Zwick}, \citenamefont {B{\"u}ttgen}, \citenamefont
  {Zorko}, \citenamefont {Tsirlin}, \citenamefont {Mendels},\ and\
  \citenamefont {Nath}}]{somesh2021}%
  \BibitemOpen
  \bibfield  {author} {\bibinfo {author} {\bibfnamefont {K.}~\bibnamefont
  {Somesh}}, \bibinfo {author} {\bibfnamefont {Y.}~\bibnamefont {Furukawa}},
  \bibinfo {author} {\bibfnamefont {G.}~\bibnamefont {Simutis}}, \bibinfo
  {author} {\bibfnamefont {F.}~\bibnamefont {Bert}}, \bibinfo {author}
  {\bibfnamefont {M.}~\bibnamefont {Prinz-Zwick}}, \bibinfo {author}
  {\bibfnamefont {N.}~\bibnamefont {B{\"u}ttgen}}, \bibinfo {author}
  {\bibfnamefont {A.}~\bibnamefont {Zorko}}, \bibinfo {author} {\bibfnamefont
  {A.~A.}\ \bibnamefont {Tsirlin}}, \bibinfo {author} {\bibfnamefont
  {P.}~\bibnamefont {Mendels}},\ and\ \bibinfo {author} {\bibfnamefont
  {R.}~\bibnamefont {Nath}},\ }\bibfield  {title} {\bibinfo {title} {Universal
  fluctuating regime in triangular chromate antiferromagnets},\ }\href@noop {}
  {\bibfield  {journal} {\bibinfo  {journal} {Phys. Rev. B}\ }\textbf {\bibinfo
  {volume} {104}},\ \bibinfo {pages} {104422} (\bibinfo {year}
  {2021})}\BibitemShut {NoStop}%
\bibitem [{\citenamefont {Ranjith}\ \emph {et~al.}(2017)\citenamefont
  {Ranjith}, \citenamefont {Brinda}, \citenamefont {Arjun}, \citenamefont
  {Hegde},\ and\ \citenamefont {Nath}}]{ranjith2017}%
  \BibitemOpen
  \bibfield  {author} {\bibinfo {author} {\bibfnamefont {K.}~\bibnamefont
  {Ranjith}}, \bibinfo {author} {\bibfnamefont {K.}~\bibnamefont {Brinda}},
  \bibinfo {author} {\bibfnamefont {U.}~\bibnamefont {Arjun}}, \bibinfo
  {author} {\bibfnamefont {N.}~\bibnamefont {Hegde}},\ and\ \bibinfo {author}
  {\bibfnamefont {R.}~\bibnamefont {Nath}},\ }\bibfield  {title} {\bibinfo
  {title} {Double phase transition in the triangular antiferromagnet
  {Ba$_3$CoTa$_2$O$_9$}},\ }\href@noop {} {\bibfield  {journal} {\bibinfo
  {journal} {J. Phys. Condens. Matter}\ }\textbf {\bibinfo {volume} {29}},\
  \bibinfo {pages} {115804} (\bibinfo {year} {2017})}\BibitemShut {NoStop}%
\bibitem [{\citenamefont {Su}\ \emph {et~al.}(2023)\citenamefont {Su},
  \citenamefont {Zeng}, \citenamefont {Sun}, \citenamefont {Sheptyakov},
  \citenamefont {Chen}, \citenamefont {Sheng}, \citenamefont {Li},\ and\
  \citenamefont {Jin}}]{Su2023}%
  \BibitemOpen
  \bibfield  {author} {\bibinfo {author} {\bibfnamefont {C.}~\bibnamefont
  {Su}}, \bibinfo {author} {\bibfnamefont {X.-T.}\ \bibnamefont {Zeng}},
  \bibinfo {author} {\bibfnamefont {K.}~\bibnamefont {Sun}}, \bibinfo {author}
  {\bibfnamefont {D.}~\bibnamefont {Sheptyakov}}, \bibinfo {author}
  {\bibfnamefont {Z.}~\bibnamefont {Chen}}, \bibinfo {author} {\bibfnamefont
  {X.-L.}\ \bibnamefont {Sheng}}, \bibinfo {author} {\bibfnamefont
  {H.}~\bibnamefont {Li}},\ and\ \bibinfo {author} {\bibfnamefont
  {W.}~\bibnamefont {Jin}},\ }\bibfield  {title} {\bibinfo {title} {Type-ii
  antiferromagnetic ordering in the double perovskite oxide
  {${\mathrm{Sr}}_{2}{\mathrm{NiWO}}_{6}$}},\ }\href
  {https://doi.org/10.1103/PhysRevB.108.054416} {\bibfield  {journal} {\bibinfo
   {journal} {Phys. Rev. B}\ }\textbf {\bibinfo {volume} {108}},\ \bibinfo
  {pages} {054416} (\bibinfo {year} {2023})}\BibitemShut {NoStop}%
\bibitem [{\citenamefont {Lin}\ \emph {et~al.}(2021)\citenamefont {Lin},
  \citenamefont {Jeong}, \citenamefont {Kim}, \citenamefont {Wang},
  \citenamefont {Huang}, \citenamefont {Masuda}, \citenamefont {Asai},
  \citenamefont {Itoh}, \citenamefont {G{\"u}nther}, \citenamefont {Russina}
  \emph {et~al.}}]{lin2021field}%
  \BibitemOpen
  \bibfield  {author} {\bibinfo {author} {\bibfnamefont {G.}~\bibnamefont
  {Lin}}, \bibinfo {author} {\bibfnamefont {J.}~\bibnamefont {Jeong}}, \bibinfo
  {author} {\bibfnamefont {C.}~\bibnamefont {Kim}}, \bibinfo {author}
  {\bibfnamefont {Y.}~\bibnamefont {Wang}}, \bibinfo {author} {\bibfnamefont
  {Q.}~\bibnamefont {Huang}}, \bibinfo {author} {\bibfnamefont
  {T.}~\bibnamefont {Masuda}}, \bibinfo {author} {\bibfnamefont
  {S.}~\bibnamefont {Asai}}, \bibinfo {author} {\bibfnamefont {S.}~\bibnamefont
  {Itoh}}, \bibinfo {author} {\bibfnamefont {G.}~\bibnamefont {G{\"u}nther}},
  \bibinfo {author} {\bibfnamefont {M.}~\bibnamefont {Russina}}, \emph
  {et~al.},\ }\bibfield  {title} {\bibinfo {title} {Field-induced quantum spin
  disordered state in spin-1/2 honeycomb magnet {$\rm Na_2Co_2TeO_6$}},\
  }\href@noop {} {\bibfield  {journal} {\bibinfo  {journal} {Nat. Commun.}\
  }\textbf {\bibinfo {volume} {12}},\ \bibinfo {pages} {5559} (\bibinfo {year}
  {2021})}\BibitemShut {NoStop}%
\bibitem [{\citenamefont {Sanjeewa}\ \emph {et~al.}(2019)\citenamefont
  {Sanjeewa}, \citenamefont {Garlea}, \citenamefont {McGuire}, \citenamefont
  {McMillen},\ and\ \citenamefont {Kolis}}]{doi:10.1021/acs.inorgchem.8b03418}%
  \BibitemOpen
  \bibfield  {author} {\bibinfo {author} {\bibfnamefont {L.~D.}\ \bibnamefont
  {Sanjeewa}}, \bibinfo {author} {\bibfnamefont {V.~O.}\ \bibnamefont
  {Garlea}}, \bibinfo {author} {\bibfnamefont {M.~A.}\ \bibnamefont {McGuire}},
  \bibinfo {author} {\bibfnamefont {C.~D.}\ \bibnamefont {McMillen}},\ and\
  \bibinfo {author} {\bibfnamefont {J.~W.}\ \bibnamefont {Kolis}},\ }\bibfield
  {title} {\bibinfo {title} {Magnetic ground state crossover in a series of
  glaserite systems with triangular magnetic lattices},\ }\href
  {https://doi.org/10.1021/acs.inorgchem.8b03418} {\bibfield  {journal}
  {\bibinfo  {journal} {Inorg. Chem.}\ }\textbf {\bibinfo {volume} {58}},\
  \bibinfo {pages} {2813} (\bibinfo {year} {2019})}\BibitemShut {NoStop}%
\bibitem [{\citenamefont {Woodland}\ \emph {et~al.}(2025)\citenamefont
  {Woodland}, \citenamefont {Okuma}, \citenamefont {Stewart}, \citenamefont
  {Balz},\ and\ \citenamefont {Coldea}}]{1pvl-kzjm}%
  \BibitemOpen
  \bibfield  {author} {\bibinfo {author} {\bibfnamefont {L.}~\bibnamefont
  {Woodland}}, \bibinfo {author} {\bibfnamefont {R.}~\bibnamefont {Okuma}},
  \bibinfo {author} {\bibfnamefont {J.~R.}\ \bibnamefont {Stewart}}, \bibinfo
  {author} {\bibfnamefont {C.}~\bibnamefont {Balz}},\ and\ \bibinfo {author}
  {\bibfnamefont {R.}~\bibnamefont {Coldea}},\ }\bibfield  {title} {\bibinfo
  {title} {From continuum excitations to sharp magnons via transverse magnetic
  field in the spin-$\frac{1}{2}$ ising-like triangular lattice antiferromagnet
  {$\rm Na_2BaCo(PO_4)_2$}},\ }\href {https://doi.org/10.1103/1pvl-kzjm}
  {\bibfield  {journal} {\bibinfo  {journal} {Phys. Rev. B}\ }\textbf {\bibinfo
  {volume} {112}},\ \bibinfo {pages} {104413} (\bibinfo {year}
  {2025})}\BibitemShut {NoStop}%
\bibitem [{\citenamefont {Kajita}\ \emph {et~al.}(2024)\citenamefont {Kajita},
  \citenamefont {Nagai}, \citenamefont {Yamagishi}, \citenamefont {Kimura},
  \citenamefont {Hagihala},\ and\ \citenamefont {Kimura}}]{Kajita2024}%
  \BibitemOpen
  \bibfield  {author} {\bibinfo {author} {\bibfnamefont {Y.}~\bibnamefont
  {Kajita}}, \bibinfo {author} {\bibfnamefont {T.}~\bibnamefont {Nagai}},
  \bibinfo {author} {\bibfnamefont {S.}~\bibnamefont {Yamagishi}}, \bibinfo
  {author} {\bibfnamefont {K.}~\bibnamefont {Kimura}}, \bibinfo {author}
  {\bibfnamefont {M.}~\bibnamefont {Hagihala}},\ and\ \bibinfo {author}
  {\bibfnamefont {T.}~\bibnamefont {Kimura}},\ }\bibfield  {title} {\bibinfo
  {title} {Ferroaxial transitions in glaserite-type {$\rm Na_2BaM(PO_4)_2$}
  ({$\rm M = Mg, Mn, Co, and Ni$)}},\ }\href
  {https://doi.org/10.1021/acs.chemmater.4c01406} {\bibfield  {journal}
  {\bibinfo  {journal} {Chem. Mater.}\ }\textbf {\bibinfo {volume} {36}},\
  \bibinfo {pages} {7451} (\bibinfo {year} {2024})}\BibitemShut {NoStop}%
\end{thebibliography}%
\end{document}